%% file: main.tex
\documentclass[runningheads]{llncs}
\input packages
\usepackage{lipsum}
\input globalcomm
\input init

\begin{document}
\maketitle

\input introduction
\input synopsis

\input inherit


\bibliographystyle{plain}
\bibliography{bibdatabase}
\end{document}

%% file: packages.tex
\usepackage[algoruled, vlined, linesnumbered, titlenotnumbered, noend]{algorithm2e}
\usepackage{listings}

\lstdefinestyle{customlang}{
  belowcaptionskip=1\baselineskip,
  breaklines=true,
  xleftmargin=\parindent,
  language=C,
  showstringspaces=false,
  basicstyle=\small\ttfamily,
  keywords={CAS,fence,while,for,goto,term,do,until},
  keywordstyle=\bfseries,
  commentstyle=\itshape\color{blue},
  stringstyle=\color{orange},
  tabsize=2
}

\SetKwBlock{Let}{let}{}
\SetCommentSty{mycommfont}
\DontPrintSemicolon 
\usepackage{ltl}
\usepackage{pifont,bbding,amsmath,stmaryrd,amssymb,mathtools} 
\usepackage{xcolor}
\usepackage{pgfmath}
\usepackage{wrapfig} 
\usepackage{xspace} 
\usepackage{hyperref}
\usepackage[capitalise]{cleveref}
\usepackage{tikz}
\usetikzlibrary{calc, positioning, backgrounds, decorations.pathmorphing, shapes, calc, arrows.meta}
\usepackage{tikz-layers}
\usepackage{graphicx}
\graphicspath{{Figures/}}
\usepackage{subfigure}

\usepackage[misc]{ifsym}

\usepackage{pgfplots}

%% file: globalcomm.tex
\newcommand\compose\circ

\newcommand\undf\bot

\newcommand\addtoset\oplus

\newcommand\ordering\preceq

\newcommand\emptyword\epsilon

\newcommand\wtranspose\otriangle

\newcommand\app\bullet

\newcommand\wfilter\odot

\newcommand\thread\theta
\newcommand\pthread\phi
\newcommand\psthread\psi

\newcommand\run\rho

\newcommand\conf\gamma

\newcommand\transitionset\Delta

\newcommand\rfrelsof[1]{{\color{relcolor}{\mathtt{{RF}}}}}

\newcommand\corelsof[1]{{\color{relcolor}{\mathtt{{CO}}}}}



%% file: init.tex
\title{Unified Fairness for Weak Memory Verification}
\author{Parosh Aziz Abdulla\inst{1}\orcidID{0000-0001-6832-6611}$^\text{\Letter}$ \and Mohamed Faouzi Atig\inst{1}\orcidID{0000-0001-8229-3481
} \and Adwait Godbole\inst{2}\orcidID{0000-0001-7704-304X} \and Shankaranarayanan Krishna\inst{3}\orcidID{0000-0003-0925-398X} \and Mihir Vahanwala\inst{4}\orcidID{0009-0008-5709-899X}} 

\authorrunning{P. A. Abdulla et. al.}

\institute{Uppsala University, Uppsala, Sweden \email{\{parosh,mohamed\_faouzi.atig\}@it.uu.se} \and
University of California Berkeley, Berkeley, USA \email{adwait@berkeley.edu} \and
IIT Bombay, Mumbai, India \email{krishnas@cse.iitb.ac.in} \and
Max Planck Institute for Software Systems, Saarland Informatics Campus, Saarbrücken, Germany \email{mvahanwa@mpi-sws.org}}

\date{}

%% file: introduction.tex

\begin{abstract}
We consider the verification of $\omega$-regular linear temporal properties of concurrent programs running under weak memory semantics. 
We observe that in particular, these properties may enforce liveness clauses, whose verification in this context is seldom studied. 
The challenge lies in precluding demonic nondeterminism arising due to scheduling, as well as due to multiple possible causes of weak memory consistency. 
We systematically account for the latter with a generic operational model of programs running under weak memory semantics, which can be instantiated to a host of memory models. 
This generic model serves as the formal basis for our definitions of fairness to preclude demonic nondeterminism: we provide both language-theoretic and probabilistic versions, and prove them equivalent in the context of the verification of $\omega$-regular linear temporal properties. 
As a corollary of this proof, we obtain that under our fairness assumptions, both qualitative and quantitative verification Turing-reduce to close variants of control state reachability: a safety-verification problem.

A preliminary version of this article appeared in the proceedings of CAV 2023 \cite{originalCAV}.
\end{abstract}

\section{Introduction}
Concurrent programs consist of multiple processes performing computations and sharing access to a memory of global variables. Decomposing a program into multiple processes may be necessitated by the setting, e.g.\ a distributed (financial) database, or may helpfully separate concerns, e.g.\ a power plant controller where there are different processes to listen for sensor input, perform physics computations, render an output reading, actuate a control setting, and so on. Given their prevalence, the formal verification of concurrent programs is naturally an important challenge. 

When writing concurrent programs, it is most intuitive to assume the processes as being fully \emph{synchronized} via the shared memory, i.e.\ the execution of the program is some interleaving of the executions of the component processes, and a read from a shared variable returns the value written by the most recent write to that variable in the interleaving. The notion of Sequential Consistency (SC) \cite{LamportSC} captures this intuition, and makes programs quite amenable to proofs of correctness.

However, real-world applications prioritize not only mathematical correctness but also performance; the latter often higher because it is easier to evaluate. As one would expect, the synchronization enforced by SC comes at a high performance cost, and is foregone in practice. We say that programs run under \emph{weak memory semantics}: here, the adjective `weak' qualifies the consistency guarantees offered by the shared memory. Typically, a process' accesses to shared memory may be reordered if they are independent enough, buffers and caches may fetch reads from shared variables speculatively, and/or procrastinate the propagation of writes.

As an example, Fig.\ \ref{verysimpleprogram} illustrates weak behavior that cannot be attributed to any sequentially consistent execution, but is permitted on ARM and POWER machines. The observed reads are possible if, for instance, in the second process, the write to $x$ is reordered before the (independent) read from $y$.

\begin{figure}[h]
    \setlength{\tabcolsep}{8pt}
    \centering 
    
    \begin{tabular}{r||l}
\begin{lstlisting}[xleftmargin=3pt,style=customlang]
a = x; //1
y = 1; 
\end{lstlisting} & 
\begin{lstlisting}[xleftmargin=3pt,style=customlang]
b = y; //1
x = 1;
\end{lstlisting}
    \end{tabular}
    \caption{Load buffering in a very simple concurrent program. Assume initially $x = y = 0$.}
    \label{verysimpleprogram}
\end{figure}

As a somewhat dual example, Fig.\ \ref{storebuffering} illustrates weak behavior that cannot be attributed to any sequentially consistent execution, but is permitted on ARM, POWER, as well as x86 TSO machines, even if instructions may not be locally reordered. This is because the writes may reside in local buffers, forcing both reads to obtain their values from the globally available initialization.

\begin{figure}[h]
    \setlength{\tabcolsep}{8pt}
    \centering 
    
    \begin{tabular}{r||l}
\begin{lstlisting}[xleftmargin=3pt,style=customlang]
y = 1; 
isync;
a = x; //0
\end{lstlisting} & 
\begin{lstlisting}[xleftmargin=3pt,style=customlang]
x = 1;
isync;
b = y; //0
\end{lstlisting}
    \end{tabular}
    \caption{Store buffering. Assume initially $x = y = 0$, and that the instructions cannot be (locally) reordered.}
    \label{storebuffering}
\end{figure}

It is clear that permitting weak behavior introduces several sources of nondeterminism beyond the basic scheduling nondeterminism of SC, and thus makes the verification task significantly more complex. This is because in general, specifications can enforce not only safety (a bad event never occurs) but also liveness (a desirable event is guaranteed to occur). Proving the latter requires appropriate \emph{fairness assumptions} on the resolution of nondeterminism. These, in turn, can only be made on a model of concurrent programs that can distinguish demonic nondeterminism. These challenges are arguably why the literature on verifying liveness for programs running under weak memory semantics is relatively sparse, despite the extensive work on verifying safety: it is only recently \cite{DBLP:conf/esop/AbdullaAAGK22,strongcohdef} that we have seen efforts to verify liveness.

The burden of formal verification (beyond safety) notwithstanding, relaxing consistency requirements improves performance as well as scalability, making concurrent programming all the more viable, and consequently, formal verification all the more critical. This article is a step in the direction of developing systematic formal verification techniques for concurrent programs running under weak memory semantics. Our key ingredient is the fairness assumption that (a) declares there is no ``discernible'' pattern in the resolution of nondeterminism, and (b) restricts the extent of weak behavior, which is quantified by our generic transition-system-based (i.e.\ operational) model of such programs.

\subsection{Preliminaries: Concurrent Programs and Verification Goal}
To establish the scope of this article, a few preliminary remarks are in order. Throughout this paper, we shall assume that a concurrent program $\mathfrak{P}$ consists of a fixed finite set of \emph{processes} or \emph{threads} (usually denoted $p_0, p_1, \dots$) that execute instructions to operate over a finite data domain $\mathbb{D}$. These processes have finitely many \emph{local variables} or \emph{registers} (denoted by letters $a, b, c$) and share access to a finite global set $\mathbb{X}$ of \emph{locations} or \emph{memory addresses} or \emph{global variables}\footnote{By convention, when we simply say `variable', we mean `global variable'.} (denoted by letters $x, y, z$). Each location (likewise, register) holds a value (usually represented by a positive integer) from $\mathbb{D}$, and is assumed to be initialized to a special value $0$. 

We consider a simple C-like programming language that includes operations on $\mathbb{D}$\footnote{Elements of $\mathbb{D}$ may be interpreted as data as well as pointers. In this paper, our concern is primarily the semantics of the shared memory accesses; it suffices to abstract away the semantics of the local computations, since the domain $\mathbb{D}$ is finite.}, goto statements, conditional jumps, non-deterministic branching, reading from shared memory into a local variable, writing a constant or a local value to shared memory, instructions to synchronize across processes, and atomic combination(s) of reads from and writes to memory.

The \emph{control state} gives for each process the current position of the \emph{program counter} (pointer to the next instruction to be completely executed) and the current values of local variables. In our formulation, there are only finitely many possible control states for any given program.

We assume that programmer intent is captured by the evolution of the control state. As an example, consider the following linear temporal property that could be imposed on the observation traces of the control state of a program with processes $p_0, p_1, p_2$: ``Eventually, $p_0$ terminates ($T_0$), and subsequently, $p_1$ and $p_2$ alternate in having exclusive access to the critical section ($C_i$).'' The verification community would specify this requirement as the $\omega$-regular language $\Sigma^*T_0 \left((N^* C_1 N^* C_2)^\omega + (N^* C_2 N^* C_1)^\omega\right)$. It is such ($\omega$-regular linear temporal) specifications\footnote{We remark that we adopt model-checking as our verification paradigm of choice as it appears better suited that deductive reasoning to the level of abstraction we are working at.} that we wish to verify the evolution of control state against. We now discuss how we model concurrent systems to achieve this verification goal.

\subsection{Modeling the System}
\begin{figure}[h]
\centering
\includegraphics[width=0.6\textwidth]{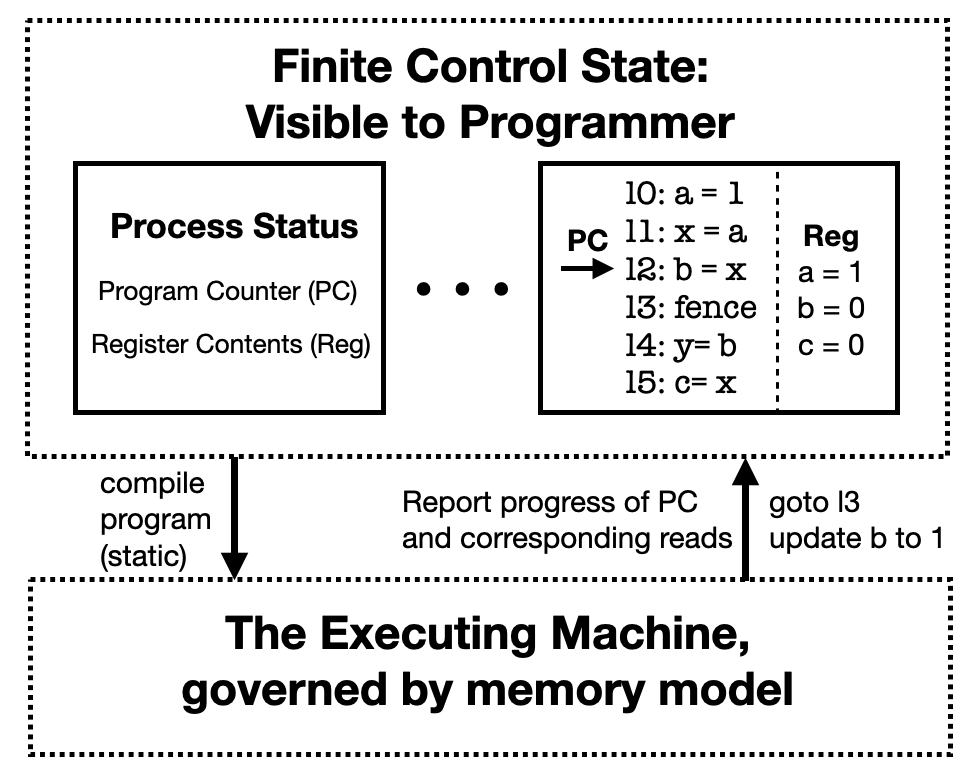}
\caption{Concurrency from a programmer's perspective}
\label{premise}
\end{figure}

In writing code to meet requirements, the programmer uses the synchronization guarantees made by the semantics of programming language. To actually execute, however, programs need to be compiled to instructions executable by the machine they run on. Naturally, the implementation of how the compiled machine code is executed must respect the semantic guarantees of the programming language. The implementation details of reads, writes, and synchronization primitives are given by a \emph{memory model}, which is an abstract description of the machine executing the program (Fig.\ \ref{premise}).

Memory models, when declared axiomatically, map an execution to a (possibly infinite) graph, and enforce its validity \emph{a posteriori} by prohibiting certain patterns in the graph. This paradigm is helpful to programmers because it conveys the semantic structure of the code that can be relied upon for correctness. Operational models, on the other hand, construct transition systems whose runs correspond to program executions: as \cite[Introduction]{armedcats} notes, they are preferable to hardware designers and verification engineers. This is because it is a more natural setting to respectively describe performance optimizations, and track various aspects of a system as it evolves with time. 

For verification in particular, $\omega$-regular specifications can be encoded as \emph{Muller automata}, which can then be composed quite naturally with a transition-system-based operational model of the execution (see Def.\ \ref{synccomp}). Satisfaction of the specification by traces of the system becomes equivalent to a connectivity property of the graph underlying the composite transition system.

We thus propose a \emph{generic operational model} (illustrated in Fig.\ \ref{operational-model}, further outlined in Sec.\ \ref{synopsismodeling}, and instantiated in Sec. \ref{instantiation}) as an abstraction of the Executing Machine of Fig.\ \ref{premise} to apply our verification techniques (sketched in Fig.\ \ref{fairness-verification}). By a generic model, we mean a ``blueprint'' with certain abstract ``parameters'', which, when instantiated, defines specific memory models. These models can have architectural origins, e.g. x86 models (RMO, PSO, TSO), the ARM memory model, the proposed model for IBM POWER, or semantic origins, e.g.\ (soundly strengthened) fragments of C/C++11 such as Strong-Release-Acquire (SRA), or even origins from distributed systems, e.g.\ Weak-RA, parallel snapshot isolation.

The notion of a generic model for weak memory is not a novelty in itself. See \cite[Section 4]{AlglaveMT14} for a generic axiomatic model, and \cite[Section 7]{AlglaveMT14} and \cite{genericmodel} for generic operational models. These advances, especially the latter, while inspirational to our techniques, do not entirely overcome our modeling challenges.

\subsection{Fairness and Verification}

\begin{figure}[h]
\centering
\includegraphics[width=0.7\textwidth]{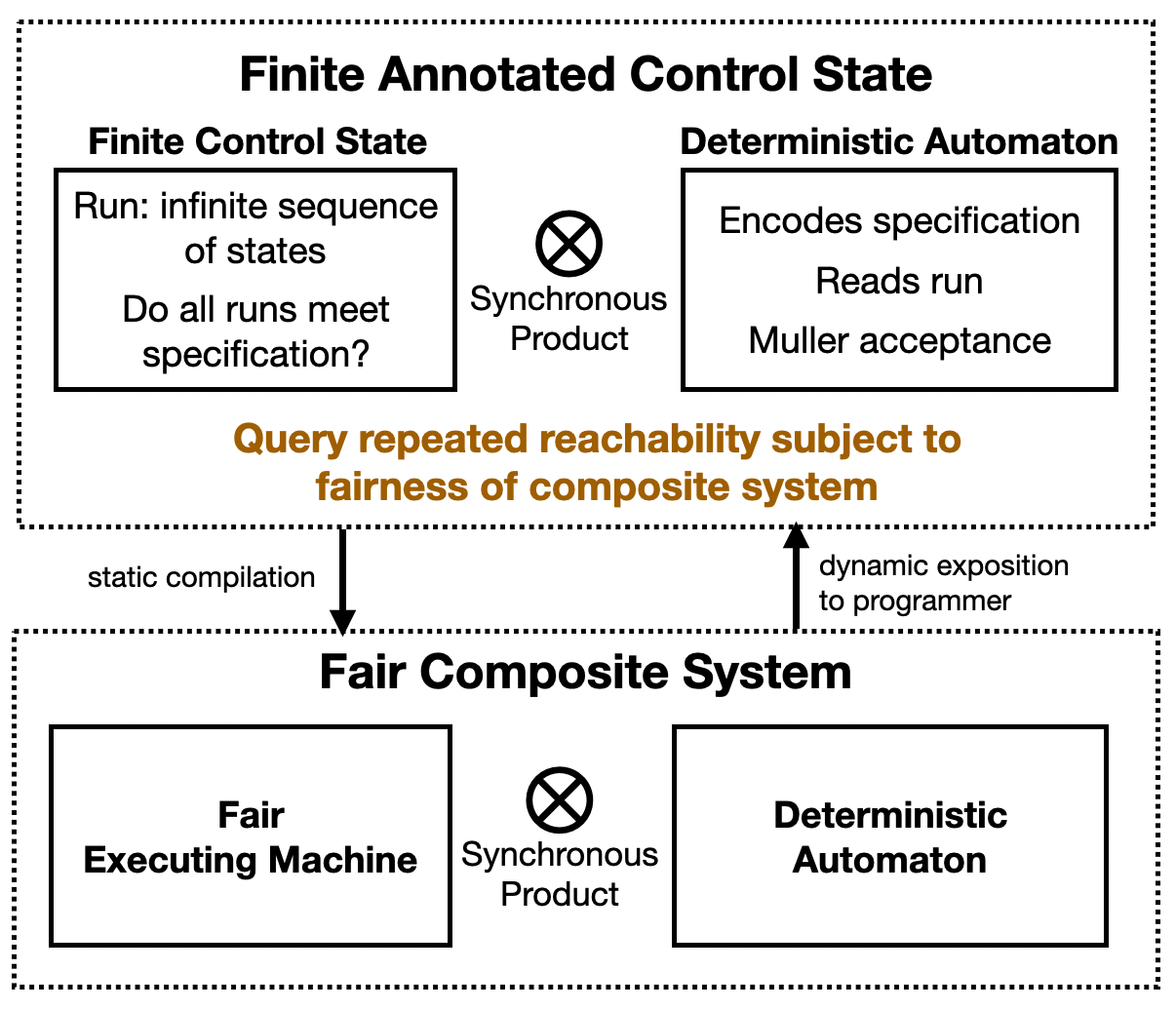}
\caption{Our blueprint to use fairness for the verification of linear temporal specifications on the evolution of control state}
\label{fairness-verification}
\end{figure}

Although generic models have been developed and proven useful, we note that the primary focus of weak memory verification, and hence modeling, has been on safety, i.e.\ the absence of fatal bugs: the modeling needed only be good enough to determine the reachability of undesirable control states. However, recall that more general specifications can also require liveness, i.e.\ the guarantee of desirable outcomes eventually occurring (e.g.\ process $p_0$ terminating). This can be done only if we make \emph{fairness assumptions} on the model to preclude the inherent non-determinism from being so demonic that favorable outcomes are denied the opportunity to occur (e.g.\ the scheduler never picks $p_0$ to run). As indicated in Fig.\ \ref{fairness-verification}, which sketches our verification approach, admitting fairness is a key feature of the modeling if verification techniques are to be applicable.

As refinements to a mathematical model of concurrent programs, it does not behoove fairness assumptions to be too ad hoc. Furthermore, as they are necessitated by practical concerns, they must also be grounded in real-world observations. Thus, in the verification of specifications more general than safety, the challenge lies not only in identifying appropriate definitions of fairness, but also in devising sufficiently perspicuous frameworks to describe memory models in a way that seamlessly admits natural fairness definitions. We illustrate how we overcome these challenges with two examples.

The program in Fig.\ \ref{transitionfairnessex} needs the second process to read the write $x = 1$ to terminate. The first process keeps alternating between writing $1$ and $2$ to $x$ until then. It would be unfair if the second process is always scheduled to read when $1$ has just been written, but never when $2$ has just been written. Transition fairness prohibits this: it enforces that if a program state is visited infinitely often, then every available transition is taken infinitely often. Technically, we need a stronger definition (see Def.\ \ref{alphafair}): if a state is visited infinitely often, then the sequence of transition choices taken does not follow any ``discernible'' pattern.
\begin{figure}[h]
    \setlength{\tabcolsep}{8pt}
    \centering 
    
    \begin{tabular}{r||l}
\begin{lstlisting}[xleftmargin=3pt,style=customlang]
while (y != 1) {
	x= 1; x = 2;
}
\end{lstlisting} & 
\begin{lstlisting}[xleftmargin=3pt,style=customlang]
while (x != 1) {}
y = 1;
\end{lstlisting}
    \end{tabular}
    \caption{Transition fairness is required to guarantee termination even under SC}
    \label{transitionfairnessex}
\end{figure}

As the program in Fig.\ \ref{memoryfairnessex} illustrates, mere transition fairness is not enough. For this program to terminate, at least one of the processes must read the other's write. However, this is not guaranteed to happen if store buffering is permitted (as in TSO) and the writes are issued more often than they are ``flushed''. This is an instance of ``overly weak'' behavior, and needs to be precluded with a notion of memory fairness. 

\begin{figure}[h]
    \setlength{\tabcolsep}{8pt}
    \centering 
    
    \begin{tabular}{r||l}
\begin{lstlisting}[xleftmargin=3pt,style=customlang]
do {x = 1;}
until (x == 2 or y == 1);
y = 1;
\end{lstlisting} & 
\begin{lstlisting}[xleftmargin=3pt,style=customlang]
do {x = 2;}
until (x == 1 or y == 1);
y = 1;
\end{lstlisting}
    \end{tabular}
    \caption{Memory fairness is required to guarantee termination}
    \label{memoryfairnessex}
\end{figure}

Fortunately, our generic framework explicitly quantifies the extent of weak behavior: in any model, it can be attributed to buffering of reads and writes, and/or to globally available writes under propagation, and is hence quantified by the total number of buffered accesses and writes under propagation. We can thus postulate definitions of memory fairness, restricting this quantity (see Defs.\ \ref{sizebounded}, \ref{repeatplain}).

In the conventional setting, we enforce the fairness of Def.\ \ref{alphafair} in conjunction with either Def.\ \ref{sizebounded} or Def.\ \ref{repeatplain}. The probabilistic analog of the second conjunction is given by Def.\ \ref{def:probmemfair}. Thm.\ \ref{equivalence} shows that all these alternate fairness assumptions are in a sense equivalent in the context of verification of $\omega$-regular linear temporal properties. As a corollary of the proof, we obtain that both qualitative and quantitative verification reduce to close variants of control state reachability: a safety-verification problem.

\subsection{Our Contributions}
This article extends \cite{originalCAV}, which appeared in the proceedings of CAV 2023, on the fronts of both modeling as well as verification. 
\begin{description}
\item [Modeling] We augment the framework introduced in \cite{originalCAV}, allowing us to capture behaviors such as speculation and racing reads (load buffering), in addition to the originally supported store buffering and delays in propagation of writes. The augmented framework is thereby capable of mirroring operational definitions of more sophisticated models such as ARMv8 \cite{armedcats} and POWER \cite[Section 7]{AlglaveMT14}. The new framework is also more interpretable, in that buffered writes are explicitly distinguished from globally available writes under propagation. We acknowledge that \cite{originalCAV} misrepresents the models of RMO \cite[Section 8 and Appendix D]{SPARCv9} and PRAM/FIFO consistency (see, e.g.\ \cite[Section 3]{fifosourceOG} and \cite[Section 3]{fifosourcesec}).

\item [Verification] Analogous to \cite{originalCAV}, we make appropriate fairness assumptions and Turing-reduce the qualitative and quantitative verification of $\omega$-regular temporal properties\footnote{The preliminary work \cite{originalCAV} handled only termination and repeated control state reachability.} to close variants of control state reachability queries. The definition of memory fairness is, in spirit, the same as that in \cite{originalCAV} because the augmented framework indeed continues to quantify the extent of weak behavior: it is straightforward to generalize the definitions of ``configuration size'' and ``plain configurations''. In order to verify more general linear temporal properties than termination and repeated reachability, however, we need a stronger language-theoretic definition of fairness. Nevertheless, we prove this definition equivalent to its probabilistic analog (Thm.\ \ref{equivalence}). 
\end{description} 

\subsubsection{Structure of the Paper}
We present the operational model in Sec.\ \ref{synopsismodeling}, briefly explain instantiations to specific models in Sec.\ \ref{instantiation}, and subsequently define both language-theoretic as well as the analogous probabilistic fairness notions and discuss the application of textbook model-checking techniques for verification in Sec.\ \ref{synopsisverification}. Our verification techniques Turing-reduce to variants of control state reachability problems for weak memory models. In Sec.\ \ref{reachabilitysubroutines}, we discuss how conventional techniques might be adapted to solve our close variants of control state reachability. Finally, we discuss related work in Sec.\ \ref{related} and offer concluding remarks in Sec.\ \ref{conclusion}.

%% file: synopsis.tex

\section{Modeling}
\label{synopsismodeling}
\begin{figure}[h]
\centering
\includegraphics[width=0.7\textwidth]{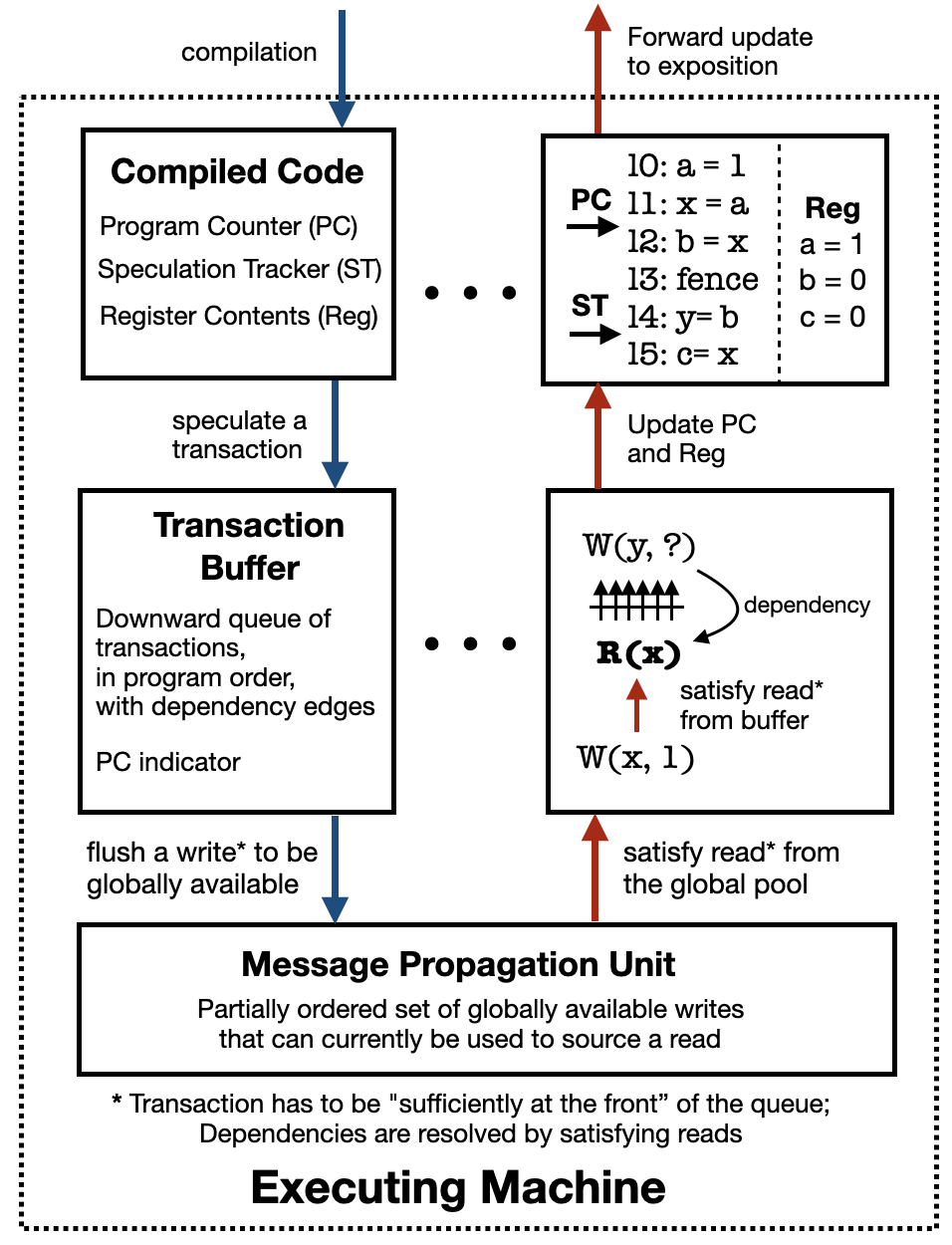}
\caption{A high-level architecture of our generic operational model, referred to as the Executing Machine in Fig.\ \ref{premise}}
\label{operational-model}
\end{figure}

In this section, we explain the generic operational model that we outline and demonstrate our generic operational model in Fig.\ \ref{operational-model}.

We remark that being a complete or authoritative source for memory models of commodity architectures such as x86, ARM, or POWER is beyond the scope of this paper, due to the sheer nuance. Nevertheless, we hope to demonstrate that our generic operational model captures them in spirit, meets the requirements  of modularity, genericity, and perspicuity, and can be proposed as an intuitive framework for architects to document their intent and designs.

The framework attributes weak behavior to two kinds of performance optimizations:
\begin{enumerate}
\item A process $p$ might locally reorder instructions that perform memory transactions, provided they are sufficiently independent. This is facilitated by the transaction buffer.
\item A write made global (i.e.\ \emph{flushed} from the buffer) by a process $p_i$ might not be immediately propagated to another process $p_j$. This delay is modeled by the message propagation unit.
\end{enumerate}

\subsection{Compiled Code}

We start understanding the framework by considering how it extends the control state: this is done in the ``compiled code'' component. To begin with, the program must be ``compiled to'' our model of the executing machine. The architecture that it is intended to be run on must also be taken into account since our framework reorders instructions rather freely by default. In our framework, compiled code mostly resembles source code, except for the insertions of instructions that ensure that the synchronization implied by programming language semantics and architectural decisions is made explicit.

\subsubsection{Compilation}
This default ability to reorder instructions, if not checked by compilation interventions, leads to gross over-approximations for two reasons. First, some reads and writes of issued in a programming language carry synchronizing semantics, e.g.\ all memory transactions issued in a language with Sequential Consistency (SC) semantics, C++ acquire loads and release stores. Moreover, some architectural models like x86 TSO (and even PSO, albeit to a lesser extent) prevent certain kinds of instruction reorderings. Our ``compilation'' captures programming language semantics and architectural rules by marking out and creating dependencies and/or inserting synchronizing instructions (see, resp.\ \emph{dependencies} and \emph{fences} in Sec.\ \ref{buffer}).

As a semantic example, acquire loads may not be overtaken by any transactions, and release stores may not overtake any transactions\footnote{ARMv8 \cite{armedcats,ARMv8} further enforces that acquire loads may not overtake release stores; to the best of our knowledge this strengthening is not corroborated by any programming language standard.}. Compilation must make these constraints explicit to the executing machine which may otherwise reorder transactions. On POWER machines, a lightweight fence is placed before release stores, and a branch and an instruction synchronization fence are placed after acquire loads \cite{cpppower1,cpppower2}. ARMv8 has added load acquire and store release instructions with similar fence semantics to support direct compilation \cite{armedcats,ARMv8}. Our compilation of such transactions to emulations of ARM or POWER on our framework mirror these fence insertions. 

As an architectural example, consider the x86 architecture, where TSO is a strengthening of RMO where reads are acquire and writes are release by default \cite[Appendix D]{SPARCv9}. Here, our compilation scheme inserts fences for \emph{architectural}, rather than semantic reasons. The distinction is subtle: if a real-world program were to be compiled to a TSO machine, release writes and acquire reads would respectively be mapped to regular writes and reads. However, we are compiling a program to run on our framework (which reorders instructions liberally) as it would on a TSO machine. Hence, we insert load-load and load-store barriers after every read, and store-store barriers after every write.

\subsubsection{State}
The state of the ``compiled code'' component of our executing machine is a slight extension of the control state that is actually exposed to the programmer. As discussed above, the compilation may insert instructions to make synchronization semantics of the program explicit. Thus, translating the program counter (PC) and register values (Reg) of this component to the exposed control state is straightforward. 

The additional feature is the \emph{speculation tracker} (ST), which steps through the code line by line, ``guessing'' the \emph{program order}, i.e.\ sequence of executed instructions. As it does so, the ST adds the memory transactions it crosses to the transaction buffer. We shall soon discuss how the buffer manages and performs the transactions: from the source code component's perspective, the buffer is responsible for driving the progress of the program counter (PC) and the attendant updates of register values (Reg), and also resetting the ST when branch misprediction is discovered. This progress in PC and Reg is then immediately conveyed verbatim to the Finite Control State exposed to the programmer.

The right of Fig.\ \ref{operational-model} illustrates an example: the ST indicates that transactions up until writing to location $y$ from the register $b$ have been \emph{speculated}; the PC indicates that the next instruction is a read from the location $x$ that will update the register $b$; finally, consistent with the progress of the PC, Reg indicates that $a=1, b=0, c=0$. At the next step, the transaction buffer can move the PC forward and convey progress along with the update of register $b$ to $1$.

There are details such as the \emph{fence} instruction, and the speculation of an \emph{unresolved} write W(y, ?) to location $y$ that our explanation has not yet specified. In order to do so, we need to understand the transaction buffer.

\subsection{Transaction Buffer}
\label{buffer}

The transaction buffer maintains a loose downward program-ordered ``queue'' of memory transactions and indicates the position of the program counter (PC). Its structure captures the idea that although the PC steps through the instructions in program order, speculation affords more flexibility in performing them: the resolution of data, address, and control dependencies can be optimized, values that reads need to return can be loaded eagerly from shared memory, and similarly writes can be stored to shared memory at leisure. In order to be \emph{satisfied} (resp. \emph{flushed}), a read (resp.\ write) transaction can ``jump the queue''. However, the overtaking, i.e.\ passing an unsatisfied read or an unflushed write, is constrained due to requirements enforced by \emph{processor self-consistency}, \emph{dependencies}, and \emph{fences}. Before discussing these terms, we explain the working of the transaction buffer through an example.

\subsubsection{Example}
We return to our running example in the right of Fig.\ \ref{operational-model}. The buffer may flush the write to $x$, or satisfy the read from $x$ using the buffered write preceding it in program order. There are two reasons why the write to $y$ may not be flushed: the first reason is that the value to be written is unresolved, and depends on the read from $x$ into register $b$; the second reason is that this write must cross a fence which prevents reordering the first three instructions with the last two.

In our example, we take the next transition to satisfy the read from $x$. This resolves the dependent write W(y, ?). In general, reads can also resolve address and control dependencies: the latter can result in (wrongly) speculated transactions getting discarded. The buffer can then choose to move the PC past the read R(x): when it does so, it notifies the source code component of the changes in PC and Reg, and also of resets in ST, if the read revealed a branch was mispredicted.

The transaction buffer tracks speculated instructions, manages the satisfaction of reads and the flushing of writes, and drives the PC. Transactions in the buffer are either \emph{active} or \emph{passive}. In the sequel, ``overtaking'' only considers jumping the active transactions in the queue.

Reads are active until they are satisfied, writes are active until they are flushed. Passive transactions are removed from the buffer upon being crossed by the PC. The PC can cross a read only if it is passive. If a write is flushed after being crossed by the PC, then it is directly removed from the buffer. Transactions that are found to be part of a mispredicted branch are immediately removed from the buffer, regardless of whether they are active. 

Observe that the buffer continues to hold satisfied reads precisely until the PC steps across them (however, a transaction passing a satisfied read on its way to the front of the queue is \emph{not} considered an overtake), but a write may await flushing even after being crossed by the PC. We also note that: (1) all transactions ahead of the one indicated by the PC must be writes; (2) the buffer holds all transactions from the PC to the ST (but some of them may be passive).

\subsubsection{Processor Self-Consistency}
Processor self-consistency refers to basic coherence requirements: (a) writes by the same process to the same location may never race; (b) a read may never have the chance to be {satisfied} (i.e.\ take its value from) by a write that comes later in program order; (c) if a read is satisfied by a write made by the same process, it must have chosen the most recent write to that location before it in program order. 

Furthermore, prohibiting reads by the same process to the same location from racing gives the stronger guarantee of SC-per-location. Several models forbid such load-load hazards that may be helpful optimizations: as notable exceptions, they were allowed by the RMO model of SPARC \cite{SPARCv9}, and pre-POWER 4 machines \cite{power4}. As \cite[Section 4.8]{AlglaveMT14} acknowledges, declaring full SC-per-location rather than processor self-consistency as an axiom may be perceived as controversial. We prefer to not enforce SC-per-location (since we explicitly discuss RMO), but instead give two types of reads, \emph{racing}, and \emph{hazard-free}, distinguished by whether they may be overtaken by reads to the same location.

\subsubsection{Dependencies}
Dependencies capture the possibility that whether, where, and what a transaction will contribute to the shared memory is determined by the outcomes of instructions preceding it in program order. The above are referred to as control, address, and data dependencies respectively. In the example in the right of Fig.\ \ref{operational-model} the write to $y$ has an unresolved data dependency on the read from $x$, which is yet to be satisfied. Satisfying the read will resolve the dependency, and \emph{determine} the write.

Dependencies directly constrain the overtaking in a buffer. For a read, address dependencies must be resolved before the buffer considers it eligible to be satisfied, either by a preceding local resolved write, or a globally available write from the propagation unit. For a write, \emph{all} dependencies must be resolved before the buffer considers it eligible to be flushed to the message propagation unit. In an operational model of the buffer, dependencies restrict the racing indirectly too: a read (resp.\ write) may not be satisfied (resp.\ flushed) if in doing so, it could violate processor self-consistency by overtaking an unresolved transaction.

\subsubsection{Fences}
Fences are the programmers' (and compilers') tools to explicitly impose synchronization within and across processes through additional constraints on the order in which transactions are executed, and on the order in which writes propagate. We shall explicitly consider only a few common types of fences: 
\begin{description}
\item [full fence] waits for preceding transactions to exit the buffer to be flushed; succeeding transactions must wait for the fence to be flushed before they can be satisfied or flushed (assumed universally available),
\item [lightweight synchronization] can never be overtaken by writes, can only be overtaken by succeeding reads when all preceding reads have been satisfied, can only be flushed when all preceding writes have been flushed (part of POWER instruction set),
\item [instruction synchronization] successors may only be satisfied or flushed after the PC crosses the fence; the fence itself is not flushed like a write but merely exits the buffer like a read (part of POWER instruction set),
\item [load acquire] a read that no transaction may overtake (part of ARMv8 instruction set),
\item [store release] a write that may not overtake any transaction and may not be overtaken by a load acquire (part of ARMv8 instruction set), 
\item [membar] combinations of barriers that prevent succeeding reads and/or writes from overtaking preceding read and/or writes (part of x86 instruction sets).
\end{description}

Fences are \emph{cumulative} if they further enforce constraints on the order in which writes are propagated \footnote{See Sec.\ \ref{propagationunit} for terminology used in this paragraph.}. Cumulativity is ensured by definition in multi-copy-atomic models like ARMv8 and the x86 family. In non-multi-copy-atomic models, cumulative fences like the full fence and lightweight synchronization are flushed to memory like writes, non-cumulative fences such as instruction synchronization leave the buffer like reads upon being crossed by the PC.

Other varied nuances (such as input/output barriers) are beyond the scope of this article; for our purposes, we shall abstractly consider other fences as barriers that prevent certain transactions from racing ahead, and may either be flushed from the buffer to enforce cumulativity, or be removed like a read. Crucially, a read cannot be satisfied, or a write cannot be flushed, if there is a fence preventing it from ``overtaking its way to the front'' of the queue.

\begin{quote}
The transaction buffer may pick a read to be satisfied (resp.\ a write to be flushed) only if doing so is guaranteed to preserve processor self-consistency, is well-determined by dependencies, and does not violate fences. 
\end{quote}

We observe that some fences have a \emph{clogging} effect on the buffer. When fences such as the full fence, instruction synchronization (isync), load acquire, or a combination of load-load and load-store membar are issued, it is impossible for succeeding transactions to be satisfied or flushed until the fence leaves the buffer. We may thus assume that instructions succeeding a clogging fence are not speculated until the fence leaves the buffer.

In particular, if we make the ARMv8 assumption that acquire loads may not overtake release stores, then all instructions are clogging in a program where all reads are acquire and all writes are release. There is no non-trivial speculation, and weak behavior can be completely attributed to the message propagation unit: this is the case for causal models like WRA, RA, and SRA. \footnote{Technically, RA and related models use message propagation axioms that make speculation redundant even without the ARMv8 non-overtaking assumption: essentially, even the relaxed loads are hazard-free, and the propagation unit subsumes store buffering. As for ARMv8, it uses the trivial message propagation unit, and hence a program entirely composed of release-acquire accesses takes Sequentially Consistent semantics, as the \href{https://developer.arm.com/documentation/102336/0100/Load-Acquire-and-Store-Release-instructions}{official documentation} acknowledges.}

\subsubsection{Handling Atomic Read-Modify-Writes (RMW)}
Some instructions, e.g. CAS, FADD, have semantics that atomically combine reads and writes. For simplicity, we assume that they have release and acquire semantics: they have the synchronization effect of a full fence, and must be satisfied directly from memory.

\subsection{Message Propagation Unit}
\label{propagationunit}
We shall now briefly discuss the mechanism underlying the satisfaction of reads. We have seen that reads may (in fact, often \emph{must}) be satisfied by a program-order preceding write, if available in the buffer. However, if there is no such write in the buffer, a read by process $p$ must be satisfied by a write from the message propagation unit that has been \emph{propagated} to it. 

At any given point in the execution of a program, there is a finite set of \emph{non-redundant} writes that are globally available to satisfy reads. For \emph{multi-copy atomic} models such as TSO, PSO, RMO \cite{SPARCv9}, and ARMv8 \cite{armedcats}, there is only one globally available write per location, which is propagated to all processes. In these models, the weak behavior is entirely attributed to the transaction buffer. When a write to location $x$ is flushed to the propagation unit, it immediately replaces the existing write.

More generally, this set may be unbounded: writes flushed from the transaction buffer need not be propagated to all other processes instantaneously. It does, however have some causal structure: propagation of writes to a process $p$ makes writes ``causally before'' them redundant to $p$. As a basic instance, recall that writes to the same location by the same process may never race. Thus, if process $p_0$ writes $x = 1$ followed by $x = 2$, and process $p_1$ uses the latter to satisfy a read, then in subsequently satisfied reads, $p_1$ cannot use the first write as a source, because it has been rendered redundant.

Memory models and the accompanying fence semantics can create further causal dependencies. Fences often have notions of \emph{cumulativity}, e.g.\ if process $p_0$ writes $x = 1$, issues a cumulative fence, and then writes $y = 1$, then the propagation of $y = 1$ to process $p_1$ implies the propagation of $x = 1$ too. Such cumulativity achieves synchronization \emph{across} processes: observing a write a thread made after a cumulative fence also informs the reading thread of knowledge the writer had at the time the fence was issued. The above example is the \emph{message passing} idiom, which is a characteristic of \emph{causally consistent} memory models such as Release-Acquire (RA) and related models SRA, WRA (see, e.g.\ \cite[Sections 3-4]{LahavXRA} for an exposition). 

Even in the presence of relaxed memory accesses like those allowed by POWER, the very semantics of acquire reads and release writes declares the cumulative synchronization described above. The standard compilation scheme of C/C++ to POWER \cite{cpppower1,cpppower2} inserts lightweight synchronization before release writes, and fake control branches followed by instruction synchronization after acquire reads. If all reads are acquire and all writes are release and the compilation proceeds as above, then \cite{LahavGV16} shows that the POWER model is equivalent to SRA.

We adapt the techniques of \cite{originalCAV} to construct the propagation unit. The unified framework of \cite{originalCAV} was a graph-based structure that captured several memory models. It recorded write messages as nodes and their dependencies as edges, and deleted nodes once their corresponding writes were rendered redundant to all processes. In this paper, our framework assumes that all the dependencies combine to form a \emph{partial order} that the propagation must respect, i.e.\ that the graph of \cite{originalCAV} is acyclic. 

A partially ordered propagation unit sacrifices the ability to capture RA and other fragments of the C++ memory model. However, it does capture the memory models of architectures that programs are eventually compiled to, e.g.\ POWER \cite[Propagation axiom, Fig.\ 18]{AlglaveMT14}. Rather than only programming language semantics, it is the architecture's implementation of the semantics that governs executions. It is the latter we focus on: our framework can capture SRA, suggested in \cite[Section 4.7]{AlglaveMT14} as a means for RA to fit the axiomatic framework developed for POWER, and subsequently formally proposed and studied in \cite{LahavGV16}.

\subsubsection{The partially ordered structure}
The message propagation unit maintains a partially ordered collection of \emph{write messages} and \emph{fence declarations}. The partial order $<$ is read ``ordered before''; if $u < v$, we say that $u$ is a predecessor of $v$ and that $v$ is a successor of $u$; by the \emph{downward closure} of a subset $S$ of a partial order, we mean the set $T = \{u: u \le v \text{ for some }v \in S\}$. A set that is equal to its downward closure is called \emph{downward closed}. Dually, we can also define the \emph{upward closure} of $S$, and an \emph{upward closed} set.

Each fence declaration records the identity of the process that issued it. Each write message records:
\begin{itemize}
\item the variable written to, and the value written;
\item the identity of the process that wrote it; \footnote{This, like the first component, is static and can only take finitely many values. For technical convenience, we may choose to not record this component explicitly, but instead assume that the program can be modified to incorporate this information in the value being written.}
\item a flag to indicate whether the write can be used as the source of an RMW (a write can be successfully ``overwritten'' by an RMW at most once);
\item the set of processes that have seen the message, either by writing it themselves, or by using it to satisfy a read;
\item the set of processes for which it is redundant.
\end{itemize}

\begin{proposition}[Invariants]
\label{invariants}
The message propagation unit maintains the following properties of the partial order.
\begin{description}
\item [Enabled Read] For every process $p$ and location $x$, there is at least one write $v$ to $x$ that can be used as the source of an RMW and is not redundant to $p$.
\item [Per-Location Coherence] For every location $x$, the set of writes to $x$ is totally ordered.
\item [Causal Propagation] For any location $x$, the set of messages to $x$ that are redundant to any process $p$ is always downward closed.
\item [Coherent Observation] For any process $p$ and messages $u, v$ to variable $x$ that are not redundant to $p$, if $v$ is seen by $p$, then it cannot be that $u < v$.
\item [Atomicity] If messages $v_0 < v_1 < \dots < v_n$ are writes to location $x$ such that for all $i < n$, $v_{i+1}$ is an RMW overwriting $v_i$, then for all writes $u$ to $x$ distinct from $v_0, \dots, v_n$, if $u < v_n$ then $u < v_0$, and dually if $v_0 < u$ then $v_n < u$.
\item [No Garbage] For every write message $v$, there is at least one process $p$ such that $v$ is not redundant to $p$; for every fence declaration $f$, there is at least one write message such that $v < f$.
\end{description}
\end{proposition}

Most notable non-multi-copy-atomic memory models (POWER, SRA) enforce per-location coherence (often referred to as modification order in C++ parlance). Furthermore, maintaining the atomicity invariant is trivial, as incoming writes are always placed at the end of the total order.

In order to discuss models like WRA and PRAM/FIFO Consistency, which are of interest as models for distributed systems, we weaken per-location coherence to po-loc. We include FIFO/PRAM consistency only for completeness' sake to rectify the misrepresentation in \cite{originalCAV}, and do not treat RMWs in FIFO. For WRA, we postulate weak atomicity.
\begin{description} 
\item [po-loc] For every process $p$ and location $x$, the writes made by $p$ to $x$ are totally ordered, and the total order is the same as their program order.
\item [Weak Atomicity] Two RMWs may not read from the same write.
\end{description}

The maintenance of these invariants are straightforward to adapt from that of the ones we discuss in this section. We refer the reader to Sec.\ \ref{instantiation} for further discussion on instantiations.

\subsubsection{Read}
A read from location $x$ by process $p$ is satisfied as follows:
\begin{description}
\item [Satisfy] Choose a write message $v_0$ to $x$ that is not redundant to $p$ and seen by $p$, and use its value to satisfy the read.
\item [Join to accumulate] Mark the following set of messages as redundant to $p$:\\ $\{u: \exists v. ~ (u \text{ and } v \text{ write to the same variable}) \land (u < v \le v_0) \}$.
\item [Garbage collection] Delete all write messages that become redundant to all processes, and then delete all fences that have no predecessor.
\end{description}

The join step from above ensures that the cumulative propagation effects enforced by the memory model and fences are conveyed to the reading process upon making an observation: if, by way of synchronization, the writer of $v_0$ were then ``aware'' of a write $v$ to variable $y$, then a write $u$ to $y$ that is ordered before $v$ must necessarily become redundant to a process $p$ that reads $v_0$. 

We present a closely related transition.
\subsubsection{Silent Update}
This transition chooses a process $p$, location $x$, and a message $v$ to location $x$, which is currently not redundant to $p$. The message $v$ is marked as seen by $p$.\footnote{Alternately, the update step can also choose to mark this message as redundant to $p$ if doing so does not violate the Enabled Read invariant.} In order to preserve the coherent observation invariant, all messages $u$ to $x$ such that $u < v$ are marked redundant to $p$. Garbage collection then deletes all writes that are redundant to all processes, and subsequently all fences that have no predecessor.

\subsubsection{Write}
We now describe the insertion of a fresh write by a process $p$ to location $x$ into the partial order. The insertion must respect the rules of the governing memory model, and also maintain the framework invariants of propagation order, location coherence, and atomicity. Of the five components of a write message, the first two are obvious. The incoming message is initially eligible to be a source of an RMW, and is \emph{seen} only by the writing thread. The set of processes for which this message is redundant is empty at the time of insertion.

\begin{description}
\item [Report accumulated observations] The latest cumulative fence issued by $p$, if present, is added as a predecessor of the write.
\item [Place as maximal element] The incoming write is ordered after all the existing writes to $x$. 
\item [Preserve coherent observation] Mark all writes $u$ to $x$ such that $u < v$ as redundant to $p$.
\item [Garbage collection] Delete all write messages that become redundant to all processes, and then delete all fences that have no predecessor.
\end{description}

\subsubsection{Fence}
When a process $p$ issues a cumulative fence, it is added as a fresh maximal element in the partial order. The set of predecessors of the newly inserted element is the downward closure of the set which is the union of: 
\begin{itemize}
\item the set of write messages seen by $p$.
\item the set of fences issued by $p$ 
\end{itemize}

\subsubsection{RMW}
Recall that we assume RMW operations to have release-acquire semantics. Thus, we atomically execute a read, issue a fence, execute a write, and mark the message read from as ineligible to source an RMW. Note that we necessarily read from the write that is maximal in the total coherence order for the location.

\subsubsection{Strong (Full) Fence}
A strong (full) fence is implemented as an RMW to an otherwise unused location.

\input framework

\section{Fairness for Verification}
\label{synopsisverification}
Equipped with an understanding of our generic operational model of weak memory, we are ready to resume discussing our verification objective. We would like to verify $\omega$-regular temporal specifications on the evolution of the control state such as, ``Eventually, $p_0$ terminates, and subsequently, $p_1$ and $p_2$ alternate in having exclusive access to the critical section.''

\subsection{Model Checking}
The standard textbook method of verifying a transition system satisfies an $\omega$-regular property\footnote{We assume all our properties to be \emph{stutter-insensitive}, because the underlying system has silent transitions (flushes and updates) that are abstract to the programmer. Stutter insensitivity means that only the sequence of distinct control states is relevant, e.g.\ $aaatbbtaatbbbt\dots$ is regarded equivalent to $atbtatbt\dots$ by the property.} is to construct its \emph{synchronous product} with a finite automaton, and check repeated reachability of states in the product transition system. This is the basis of our approach as well, which we illustrate in Fig.\ \ref{fairness-verification}. The alphabet of the automaton is the finite set of control states of the program (all possible configurations of program counters and register values). We remark that we use a \emph{deterministic} automaton (with Muller acceptance condition to get the expressive power of $\omega$-regular languages) to ensure that the fairness assumptions on the composite, or annotated, system only have one source of non-determinism to constrain: that arising from the original system.

More formally, we let $\Gamma$ be the set of states of the (instantiated) executing machine, and $C$ be the finite set of control states of the concurrent program $\mathfrak{P}$.\footnote{Our definition of $\Gamma$ subsumes $C$, but as motivated, we decide to make the exposition of the control state to the verification techniques explicit.} The deterministic finite automaton $\mathcal{A}$ has a set $Q$ of states, starts in initial state $q_0$, reads the alphabet $C$, has a transition function $\Delta: Q \times C \rightarrow Q$, and the Muller acceptance condition, given by $\mathcal{F} \subseteq 2^Q$. The automaton $\mathcal{A}$ accepts a trace $\alpha \in C^\omega$ if and only if $F \in \mathcal{F}$, where $F$ is the set of states visited infinitely often by the run of $\mathcal{A}$ on $\alpha$.

\begin{definition}[Synchronous Composition]
\label{synccomp}
Let a given program $\mathfrak{P}$ and memory model induce a state space $C \times \Gamma$, and let $\mathcal{A}$ be a deterministic Muller automaton over the alphabet $C$. The state space of the annotated system is defined as $Q \times C \times \Gamma$. The initial state is $(\Delta(q_0, c_{\text{init}}), c_{\text{init}}, \gamma_{\text{init}})$. For every transition $(c, \gamma)$ to $(c', \gamma')$ in the original system, and every state $q \in Q$, we define the transition $(q, c, \gamma)$ to $(\Delta(q, c'), c', \gamma')$ in the annotated system.
\end{definition}

Computing the possible sets of infinitely-often-visited automaton states in the synchronous product is central to the verification techniques. However, it is important to specify what ``possible'' means for the results to be meaningful: can the space of possibilities contain \emph{any} execution trace?

Specifications that have a liveness clause, e.g.\ ``Eventually $p_0$ terminates...'' have a trivial, albeit unreasonable infinite runs that violate them, e.g.\ a run that never schedules $p_0$. Here, the scheduler is being adversarially unfair.

As a more involved, but trivially constructed counterexample to the specification, consider a situation where $p_0$ needs to read flags set by $p_1$ and $p_2$ in order to terminate. This may never happen if the transaction buffers of $p_0$, $p_1$, and $p_2$ are not flushed (thereby restricting $p_0$ to reading its own writes), or if the writes flushed by $p_1$ and $p_2$ are never propagated to $p_0$. In these cases, the memory subsystem, which is supposed to serve as a medium of communication between the processes, is clearly doing a poor job: the weak memory is arguably too weak to be fair.

We thus need to restrict the set of ``possible'' runs to those that satisfy some notion of \emph{transition fairness}, and do not exhibit unfettered weak behavior. 

\subsection{Transition Fairness}
\label{transfair}
The basic idea of transition fairness \cite{DBLP:conf/lpar/AminofBK04} is that if a non-deterministic choice is presented infinitely often, then each of the available options must also be taken infinitely often.

\begin{definition}[Transition Fairness]
A run $\alpha \in \Gamma^\omega$ of a transition system with state space $\Gamma$ is transition-fair, if: for every state $\gamma \in \Gamma$ that is visited infinitely often, each transition enabled from $\gamma$ is also taken infinitely often.
\end{definition}

Transition fairness is often assumed in solving repeated reachability queries. Our verification task is to determine the repeated reachability of states in the composite system, which are of the form $(q, \_, \_) \in Q \times C \times \Gamma$, and hence, we would like the composite system to be transition fair. However, we can only declare fairness assumptions on the executing machine.

Unfortunately, there is a caveat: a transition fair run of the original system may not correspond to a transition fair run of the composite/annotated system. Each available choice of transition from state $(c, \_)$ of the original system being taken infinitely often does not imply that each available choice is also taken infinitely often from the composite state $(q, c, \_)$. If $(c, \_)$ had successors $(c_0, \_)$ and $(c_1, \_)$ that it chose alternately, the run would be transition fair. However, an automaton can track the parity of the number of times $c$ is visited, through states $q_0$ and $q_1$. In the composite system, the transition to $c_1$ would be enabled infinitely often from the repeatedly visited $(q_0, c, \_)$, but never taken.

Although the run of the composite system being transition fair is not logically guaranteed, it would nevertheless be so with probability $1$ were the non-determinism to be resolved stochastically. Thus, this caveat is inapplicable in the probabilistic setting because the choice is resolved in a \emph{memoryless} manner. 

To guarantee that the annotated system is transition fair, the resolution of nondeterminism in the original system indeed needs to be memoryless. Our proposed fairness notion on the original system is inspired by the $\alpha$-fairness identified by Pnueli and Zuck \cite[Section 5]{pnuelialpha}.

\begin{definition}[$\alpha$-Transition Fairness]
\label{alphafair}
A run $\alpha \in \Gamma^\omega$ of a transition system state space $\Gamma$ is $\alpha$-transition fair, if for every deterministic finite automaton with states $Q$, the corresponding run $\beta \in (Q \times \Gamma)^\omega$ in synchronous composition (Definition \ref{synccomp}) with state space $Q \times \Gamma$ is transition-fair.
\end{definition}

We enforce $\alpha$-transition fairness on the executing machine. Although it seems quite restrictive from a language-theoretically perspective, the above definition merely declares that there is no automatic ``pattern'' to the scheduler's decision making. This is a reasonable abstraction for verification techniques: if we chose to abstract the choice of transitions as probabilistic, then runs would be $\alpha$-transition fair with probability $1$.\footnote{The set of runs that are unfair when composed with some fixed $\mathcal{A}$ has measure $0$; the set of $\alpha$-transition unfair runs is a countable union over all automata, and hence also has measure $0$.}

\subsection{Memory Fairness}
Transition fairness, on its own, unfortunately fails to preclude unfairly weak behavior, because the unbounded buffers of our model enable the condition to be vacuously satisfied: A run could simply choose to never flush its buffers, thus visiting a new state at every time-step.

We need a notion of memory fairness to supplement transition fairness. It is here that the perspicuity of our framework comes to the fore. Recall that we explicitly attribute weak behavior to transaction buffering and delays in the propagation unit. By construction, these structures only keep track of \emph{relevant} information: the buffer discards mispredicted transactions, retains writes only till they are flushed, and retains reads only till their values are returned to the control state; the propagation unit garbage-collects writes that have been rendered redundant to each process. Thus, our framework \emph{quantifies} the extent of weak behavior through its state: larger the total size of the memory subsystem (buffers and the set of the propagation unit), weaker the behavior.

\begin{definition}[Configuration Size]
The size of state $\gamma \in \Gamma$ of the executing machine is defined as the total number of transactions (both active and passive) in the transaction buffers, plus the total number of messages in the propagation unit.
\end{definition}

The memory fairness definitions we propose can be summarized as, ``Restrict the growth of the configuration size.'' We note that overly weak behavior is rarely observed in practice \cite{ElverN14,aros-isca15}. We explaining this by noting that the physical hardware that implements the caching and buffering is inherently finite-state, and flushed regularly. Informally, the finite footprint of the system architecture (eg. micro-architecture) implies a bound, albeit hard to compute, on the size of the configurations of the executing machine. Thus, we use the notion of configuration size to define:

\begin{definition}[Size Bounded Executions]
\label{sizebounded}
    An execution $\gamma_0\gamma_1\dots \in \Gamma^\omega$ is said to be size bounded, if there exists an $N$ such that for all $n \in \mathbb{N}$, $\gamma_n$ has size at most $N$. If this $N$ is specified, we refer to the execution as $N$-bounded.
\end{definition}

The notion of size-boundedness is an excellent start to the objective of restricting the growth of the configuration size. However, it is not immediately clear how this fairness condition integrates with verification techniques. If $N$ is known but large, the ensuing finite-state model checking might still be infeasible. This is also the reason why we need techniques more general than model checking parametrized by $N$, if the bound is unknown.

We observe that if we enforce both size boundedness and $\alpha$-transition fairness, then arbitrarily long sequences of flushes and silent updates will be taken infinitely often. This means that configurations of minimal size will also be visited infinitely often.

\begin{definition}[Plain Configurations]
\label{plainconfig}
A configuration is called plain if:
\begin{itemize}
\item The transaction buffer is empty, i.e.\ the speculation tracker coincides with the program counter, and no writes await flushing.
\item The message propagation unit has exactly one message per location.
\item All messages are seen by all processes.
\end{itemize}
\end{definition}
Intuitively, plain configurations epitomize SC-like behavior as they are states where the processes are fully synchronized (see Sec.\ \ref{plainprop} for further discussion). The repeated reachability of plain configurations is the ``limit'' of $N$-bounded fairness as $N$ diverges to infinity. We use this intuition for our key definition of memory fairness.

\begin{definition}[Repeatedly Plain Executions]
\label{repeatplain}
    An execution $\gamma_0\gamma_1 \dots \in \Gamma^\omega$ is said to be repeatedly plain, if $\gamma_n$ is a plain configuration for infinitely many $n$.
\end{definition}

In the classical, i.e.\ non-probabilistic setting, we enforce either size-bounded $\alpha$-transition fairness, or repeatedly plain $\alpha$-transition fairness on the executing machine.

\subsection{Probabilistic Memory Fairness}
Quantitative verification is motivated by the argument that conventional logic does not inherently capture the nuance of real-world systems and expectations. For example, a 1\% bound on the probability of failure is much more indicative than a declaration that a system might fail. Even in our setting, viewing the transition system from a quantitative (probabilistic) perspective \emph{implicitly} produced a much more intuitive definition of a stronger transition fairness (Sec.\ \ref{transfair}) than an explicit language-theoretic one. 

One can consider the quantitative verification of programs running under weak memory too. The executing machine is naturally viewed as an infinite-state transition system, and transitions can be assigned probabilities, thus inducing a \emph{Markov chain}. Typical questions on the Markov-chain executing machine include:
\begin{enumerate}
\item Is an $\omega$-regular property almost surely satisfied (i.e.\ with probability $1$)?
\item Approximately with what probability is an $\omega$-regular property satisfied?
\end{enumerate}

The above include liveness properties, and we indeed need notions of fairness analogous to the conventional setting. We already have $\alpha$-transition fairness, and so we need to assign transition probabilities in a way that makes the system memory-fair too.

\begin{definition}[Probabilistic Memory Fairness]
    \label{def:probmemfair}
    A Markov-chain executing machine is considered to satisfy probabilistic memory fairness
    if a plain configuration is reached infinitely often with probability one.
\end{definition}

This parallel has immense utility because verifying liveness properties for a class of Markov chains called decisive 
Markov chains\footnote{In these Markov chains, for every set $S$ of states, with probability $1$, the run either eventually reaches $S$, or eventually reaches a state $\gamma$ from which $S$ is unreachable. There is $0$ probability of a run being forever indecisive about reaching $S$.} is well studied. In \cite{AbdullaHM07}, it is established that the existence of a 
\textit{finite attractor}, i.e a finite set of states $F$ that is repeatedly reached with probability $1$, is sufficient for decisiveness. The above definition asserts that the set of plain configurations is a finite attractor.

\subsection{Equivalence of Fairness Notions}
In \cite{originalCAV}, it was proven that size bounded transition fairness\footnote{The original paper always considered regular transition fairness, not the $\alpha$-strengthening.}, repeatedly plain transition fairness, and probabilistic memory fairness are equivalent with respect to the representative liveness problems of repeated control state reachability and termination. We restate the result here (the obvious analogous statement holds for termination).
\begin{theorem}[Equivalence result of \cite{originalCAV}]
\label{theorem:equivalenceOG}
There exists $N_0 \in \mathbb{N}$ such that for all $N \ge N_0$, the following are equivalent for any control state (program counters and register values) $c$:
\begin{enumerate}
\item All $N$-bounded transition fair runs visit $c$ infinitely often.
\item All repeatedly plain transition fair runs visit $c$ infinitely often.
\item $c $ is visited infinitely often under probabilistic memory fairness with probability $1$.
\end{enumerate} 
\end{theorem}

We adapt the proof to make an extended claim.
\begin{theorem}[Extended Equivalence Result]
\label{equivalence}
Let $\mathfrak{P}$ be a concurrent program with control states $C$, and let $\Gamma$ be the set of states of the executing machine. Let $\mathcal{A}$ be a deterministic automaton with states $Q$ recognizing a stutter-insensitive $\omega$-regular language with Muller acceptance given by $F \subseteq 2^Q$. There exists an $N_0$, depending on the program, executing machine, and $\mathcal{A}$, such that for all $N \ge N_0$, the following are equivalent.
\begin{enumerate}
\item $\mathcal{A}$ accepts all control state traces generated $N$-bounded $\alpha$-fair runs of the executing machine.
\item $\mathcal{A}$ accepts all control state traces generated repeatedly plain $\alpha$-fair runs of the executing machine.
\item A control state trace generated by a probabilistic memory fair executing machine is accepted by $\mathcal{A}$ with probability $1$.
\end{enumerate}
\end{theorem}
\begin{proof}
In this proof, we shall work with the synchronous product (Def.\ \ref{synccomp}) of automaton, control state, and executing machine. Our state space is therefore $Q \times C \times \Gamma$. Note that the ``choice'' of transition lies entirely with $\Gamma$: this choice drives the next value of the $C$-component, and further cascades to the $Q$-component. We shall refer to $Q \times C$ as \emph{annotated control state}.

We shall show that the possible sets of automaton states seen infinitely often in a run is the same in all cases, i.e. $S \subseteq Q$ can be the set of states visited infinitely often by some run in Case 1 iff it can be so in Case 2 iff it can be so in Case 3. All three cases are guaranteed to accept runs if and only if all possibilities of infinitely visited states $S_1, \dots, S_k \in \mathcal{F}$. We begin to observe the parallels with Thm.\ \ref{theorem:equivalenceOG}: the setting involves repeated reachability of annotated control states, and declaring $\alpha$-transition fairness on $\Gamma$ gives transition fairness of the composite system.

We define a directed \emph{connectivity graph} $\mathcal{G}(N)$, parametrized by $N$. It has a fixed finite set of vertices, one corresponding to each $(q, c, \gamma_p) \in Q \times C \times \Gamma_{\text{plain}}$. We refer to this finite restriction of the state space of the composite transition system as \emph{annotated plain configurations}. In $\mathcal{G}(N)$, we draw an edge from $(q, c, \gamma)$ to $(q', c', \gamma')$ iff $(q', c', \gamma')$ can be reached from $(q, c, \gamma)$ via configurations of size at most $N$. Each vertex $v$ of $\mathcal{G}(N)$ also records a set $R \subseteq Q$: for the vertex corresponding to $(q, c, \gamma)$, $$R(v) = \{q': (q', c', \gamma') \text{ is $N$-reachable from } (q, c, \gamma) \text{ for some } c\in C, \gamma' \in \Gamma \},$$
where $N$-reachable stands for reachable via configurations of size at most $N$.

We similarly define $\mathcal{G}(\infty)$, where there is no bound on the size of intermediate configurations in the reachability requirements to draw an edge or add an element to the $R$-set of a vertex. 
We note:
\begin{enumerate}
\item There are only finitely many possibilities for $\mathcal{G}(N)$.
\item $\mathcal{G}$ is monotone. As $N$ increases, edges can only be added to $\mathcal{G}(N)$, and states can only be added to the $R$-set of any node. Monotonicity and finiteness guarantee saturation, i.e.\ there exists an $N_0$ such that for all $N \ge N_0$, $\mathcal{G}(N) = \mathcal{G}(N_0)$.
\item Any witness of reachability is necessarily finite, hence the saturated graph is the same as $\mathcal{G}(\infty)$, i.e.\ we can define $\mathcal{G} = \mathcal{G}(N_0) = \mathcal{G}(\infty)$.
\end{enumerate}

This $\mathcal{G}$ defined in the last point will serve as our canonical object to establish the equivalence of the fairness conditions. Now, let $V_1, \dots, V_k$ be the bottom strongly connected components (BSCCs) of $\mathcal{G}$ reachable from the node $v_{\text{init}} \in \mathcal{G}$ corresponding to $(\Delta(q_{\text{init}}, c_{\text{init}}), c_{\text{init}}, \gamma_{\text{init}})$. 

Define $S_i = \cup_{v \in V_i} R(v)$. (i) We will argue that for every $S_i$, there is a non-zero probability (and hence fair runs corresponding to Cases 1 and 2 such) that the set of automaton states visited infinitely often is $S_i$. (ii) Conversely, we will also show that with probability $1$ (resp.\ given fairness of Cases 1 and 2), the set of automaton states visited infinitely often by a run is one among $S_1, \dots, S_k$. This implies that the guarantee on automaton acceptance holds iff $\{S_1, \dots, S_k\} \subseteq \mathcal{F}$.

For (i), since $V_i$ is a reachable BSCC, there is a finite path in the transition system that enters it. This serves as the prefix of the set of runs whose infinitely-visited set is $S_i$. Let the probability of such paths being taken be $\mu_i > 0$. Probabilistic memory fairness guarantees that plain configurations are visited infinitely often with probability $1$: thus a non-empty subset of annotated plain configurations in $V_i$ will be repeatedly visited with non-zero probability $\mu_i$. Markov fairness then guarantees that given that a state a visited infinitely often, all states reachable via a finite path will also be visited infinitely often with probability $1$. Thus, with probability $\mu_i$, \emph{all} annotated plain configurations of $V_i$ are visited infinitely often. Finally, we apply transition fairness to argue that with probability $\mu_i$, all automaton states accessible from $V_i$ (by definition, $S_i$) are precisely the ones visited infinitely often.

The converse (ii) follows immediately in the probabilistic setting, because Markov fairness along with memory fairness guarantees that the probabilities $\mu_i$ of landing in BSCC $V_i$, and hence visiting $S_i$ infinitely often, add up to $1$. Similarly, avoiding a BSCC forever violates the conjunction of transition and memory fairness. As we argued for (i), we can show that $\alpha$-transition-fair and memory fair runs must necessarily access all reachable automaton states in $S_i$, having entered $V_i$.

Thus, infinitely visited sets of fair runs can be any of $S_1, \dots, S_k$, and must be one of them. Automaton acceptance is guaranteed iff $\{S_1, \dots, S_k\} \subseteq \mathcal{F}$. The definitions of $S_1, \dots, S_k$ yield the same sets, independent of the fairness condition. This concludes the proof of equivalence of fairness notions.
\qed
\end{proof}

\subsection{Verification Algorithms}
We now discuss how the proof of Thm.\ \ref{equivalence} gives us algorithms for both conventional and quantitative verification. The key structure is the connectivity graph $\mathcal{G}$, whose vertices correspond to annotated plain configurations, record automaton states reachable from these configurations, and edges indicate reachability between annotated plain configurations. All our techniques Turing-reduce to the construction of the graph $\mathcal{G}$, which, as discussed in Sec.\ \ref{reachabilitysubroutines} involves queries which are slightly refined instances of control state reachability. In the rest of this section, we shall assume that $\mathcal{G}$ has been constructed.

The conventional model checking algorithm (and the almost-sure model checking algorithm in the probabilistic setting) are derived immediately from the proof of Thm.\ \ref{equivalence}. Given $\mathcal{G}$, we simply find its BSCCs $V_1, \dots, V_k$, read off the corresponding state-sets $S_1, \dots, S_k \in 2^Q$ (where $S_i = \cup_{v\in V_i}R(v)$), and check that $\{S_1, \dots, S_k\} \subseteq \mathcal{F}$.

\subsubsection{Quantitative Model Checking}
To solve the quantitative model checking problem (i.e. approximate, to arbitrary precision, the probability that automaton $\mathcal{A}$ accepts the trace of control states produced by the run), the principle is the same. We start by identifying \emph{accepting} BSCCs: $V$ is an accepting BSCC iff $S = \cup_{v\in V}R(v)$ is in $\mathcal{F}$. Let the accepting BSCCs be $U_1, \dots, U_\ell$. By the same arguments as in the proof of Thm.\ \ref{equivalence}, the probability of acceptance is equal to the probability that a run of the composite system reaches an accepting BSCC.

It now remains to approximate the probability that an accepting BSCC is reached. In order to do so, we explore the set of all run-prefixes in a breadth-first manner. The breadth-first search maintains a FIFO queue, whose elements represent run-prefixes by pairs of state and probability. The exploration maintains two aggregates, one for the probability of acceptance (reaching an accepting BSCC), one for the probability of rejection (all accepting BSCCs becoming inaccessible).  

The exploration starts from the initial state of the composite system: this is the beginning of all runs, and has probability $1$. At each step, the front of the queue is popped. If the state of this element corresponds to a node of $\mathcal{G}$ in an accepting BSCC, we add its probability to the acceptance aggregate; if the state of the element corresponds to a node of $\mathcal{G}$ from which no accepting BSCC is accessible, we add its probability to the rejection aggregate. Otherwise, we compute successor elements by taking transitions from the state with their respective probabilities, and add them to the exploration queue.

Note that the acceptance aggregate is an under-approximation for the acceptance probability, and $1$ minus the rejection aggregate is an over-approximation. Since the system is stochastic, the aggregates always add up to at most $1$. The approximations are sound by construction; we next prove that they can get arbitrarily precise.

In order to prove that the approximation scheme terminates for any required precision, we must show that the sum of the aggregates can get arbitrarily close to $1$. Markov transition fairness and memory fairness guarantee that a run will eventually contribute to either aggregate with probability $1$. Indeed, consider the set of runs which do not have any prefix that contributes to either of the aggregates. This can only be because:
\begin{itemize}
\item The run never visits an (annotated) plain configuration (violation of memory fairness)
\item The run never settles into a BSCC (violation of transition/Markov fairness)
\end{itemize}
These events occur with probability $0$, and failing to account for such non-contributing runs is inconsequential to quantitative verification. 

\section{Reachability Subroutines}
\label{reachabilitysubroutines}
We noted that our verification algorithms Turing-reduce to reachability problems to and from annotated control states and/or annotated plain configurations. These are slightly refined versions of the regularly studied control state reachability problems. In this section, we comment on how existing techniques may be adapted to solve our custom refinements. 

We first deal with the replacement of control states (as targets) and the initial configuration (as the source) by plain configurations, and show that it does not make the problem any harder. We then make an informal case for why annotations may be handled by existing decision techniques without significant adaptation. Formally, the addition of automaton-state annotations to the control state is a caveat of a much more technical nature, and it is unclear how to rigorously address it in any generality, given the nuance of different memory models.

\subsection{Plain Configurations as Checkpoints}
\label{plainprop}
Plain configurations are central to our verification techniques, because they serve as ``reachability checkpoints'' in our algorithms. We record some of their properties in this section. 

The invariants maintained by the propagation unit (Prop.\ \ref{invariants}) imply that:
\begin{itemize}
\item Each of the writes in the propagation unit may be used to source an RMW, and no write is redundant to any process (due to Enabled Read and the implementation of writes).
\item Subsequent incoming writes will be coherence-after the writes in the present configuration (due to Per-Location Coherence).
\item Subsequent release writes will accumulate all of the writes in the present configuration (from implementation of fences).
\end{itemize}

\begin{lemma}
\label{ctrltoplain}
The decision problem of plain configuration reachability reduces to that of control state reachability.
\end{lemma}
\begin{proof}
We use a common trick to transform the given program $\mathfrak{P}$ into an augmented program $\mathfrak{P}'$ as follows.
\begin{enumerate}
\item Augment the datatype to $\mathbb{D} \times \text{Procs}$, where Procs is the finite set of processes. Each value has two components, the value used in the computation of $\mathfrak{P}$, and the identity of the writing process.
\item Declare a new shared variable to serve as a counter for a turnstile barrier.
\item At program counter labels of the desired control state in $\mathfrak{P}$, add branch instructions that check whether the register values are as desired, and if so, nondeterministically jump to distinguished code added in $\mathfrak{P}'$.
\item Having made the jump in $\mathfrak{P}'$, the processes execute the distinguished code to fully synchronize via a turnstile barrier, and check whether the shared variables of $\mathfrak{P}$ hold values as desired by the plain configuration.
\end{enumerate}

The turnstile barrier works as follows: each process atomically increments the counter using an RMW instruction (with release-acquire semantics), and then using an atomic RMW, busy-waits until the value of the counter is equal to the number of processes. The process is then repeated a second time.

At this point, a plain configuration is guaranteed to have been reached. The first round of the turnstile barrier ensures all buffers are flushed, and already guarantees a plain configuration in multi-copy-atomic models. In non-multi-copy-atomic models, the last RMW of the first round will have accumulated the knowledge regarding writes to locations used in $\mathfrak{P}$ of all processes. The second round of the turnstile then synchronizes the processes as they acquire this knowledge by reading the counter.

Crucially, when restricted to the values of the variables of $\mathfrak{P}$, the attained plain configuration can be \emph{any} of the plain configurations that could have been reached without executing the RMWs, i.e.\ merely through flushes or silent updates. Finally, each process, upon crossing the second turnstile, simply reads all locations of $\mathfrak{P}$ and checks that they hold the values stipulated by the plain configuration.
\qed
\end{proof}

The above reduction only uses the strong cumulativity of RMWs and their ability to work as fences (even for WRA, see \cite[Example 3.9]{LahavXRA}). The proof works for RA and WRA as well (because the read value also identifies the writing thread, and \emph{all} processes read to check). 

The full synchronization used to enforce plain configurations in the above reduction can also be used to prove the following.

\begin{lemma}
The problem of control state reachability from a given configuration reduces to that of control state reachability (from the initial configuration).
\end{lemma}

The reduction constructs an augmented program that uses strong synchronization to write the desired values to each location, and to ensure that each process has seen these writes, and thereafter runs like the original program.

\subsection{Accounting for Annotations}

We argue that the refinement to the conventional decision problem by composing control state with the state of an automaton is mostly syntactic in nature. For memory models such as TSO, PSO (see \cite{DBLP:conf/esop/AbdullaAAGK22,DBLP:conf/popl/AtigBBM10} for expositions of the models and decidability techniques), SRA, WRA (see \cite{LahavXRA} for a comprehensive exposition and a proof of decidability), StrongCOH (the repaired C11 model restricted to relaxed accesses; see \cite[Section 2.3]{strongcohdef} for a definition, \cite[Section 5]{strongcohdecide} for proof of decidability) where control state reachability is shown decidable, the techniques naturally demarcate the control state from the memory subsystem (i.e.\ the remaining part of the executing machine), and formulate the latter as a well-structured transition system (WSTS) \cite{AbdullaBSL10}. Recombining the finite control state retains the WSTS property, and hence makes control state reachability decidable by virtue of being a coverability problem. The same techniques apply if the control state were annotated, because it still remains finite. 

Put differently, it is common to abstract the control state as a finite transition system. This automaton is highly ``decoupled'': it has at least one strongly connected component per process, that is disconnected from the rest of the system. Composing the control state with an automaton to produce the annotated control state has a ``coupling'' effect: the automaton is no longer disconnected. However, WSTS techniques seem to treat the control state as an abstract finite object: they appear to be agnostic to the structure of the transition system. We believe the techniques would often go through regardless, as long as the interface between memory and control state remains unchanged, which in our case, it does.

%% file: framework.tex

\section{Instantiating Our Framework}
\label{instantiation}
In this section, we briefly describe various memory models with the help of example \emph{litmus tests}, and thus intuit how our framework can be instantiated to specific models. Table \ref{modelsources} summarizes the sources we use for each of the models we consider.

\begin{table}[h!]
    \centering
    \begin{tabular}{|l|l|l|}
        \hline
        \textbf{Memory Models} & \textbf{Sources} & \textbf{Notes} \\
        \hline
        RMO, PSO, TSO & \cite[Section 8, Appendix D]{SPARCv9} & Official documentation of SPARCv9 \\
        \hline
        PSO, TSO & \cite{DBLP:conf/popl/AtigBBM10} & Decidability of control state reachability \\
        \hline
        TSO & \cite{DBLP:conf/esop/AbdullaAAGK22} & Probabilistic verification \\
        \hline
        ARM(v8) & \cite{armedcats,ARMv8} & Formal modeling by ARM \\
        \hline
        POWER & \cite{AlglaveMT14} & Model based on extensive testing \\
        \hline 
        SRA & \cite{LahavGV16} & Proposed to strengthen C/C++ RA \\
        \hline  
        WRA* & \cite{BouajjaniEGH17,LahavXRA} & A minimal model of causal consistency \\
        \hline
        SRA, WRA* & \cite{LahavXRA} & Decidability of control state reachability \\
        \hline
        PSI* & \cite[Section 3.1]{LahavXRA} & Parallel Snapshot Isolation \\
        \hline
        FIFO/PRAM* & \cite{fifosourceOG} and \cite[Section 3]{fifosourcesec} & FIFO Consistency \\
        \hline
    \end{tabular}
   
    \caption{A summary of sources we use for memory models. Models marked * are relevant as consistency models for distributed systems.}
    \label{modelsources}
\end{table}

\subsection{Multi-Copy-Atomic Models}
In models such as RMO, ARMv8, PSO, and TSO, weak behavior is attributed entirely to instruction reordering in the transaction buffer. The message propagation unit always consists of a single write per variable, which is rendered redundant upon a new write to that variable being flushed from a buffer. The independent reads of independent writes (IRIW) litmus test (Fig.\ \ref{iriw}) distinguishes multi-copy-atomic models from non-multi-copy-atomic ones.

\begin{figure}[h]
    \setlength{\tabcolsep}{8pt}
    \centering 
    
    \begin{tabular}{c||c||c||c}
\begin{lstlisting}[xleftmargin=3pt,style=customlang]
x = 1;
\end{lstlisting} & 
\begin{lstlisting}[xleftmargin=3pt,style=customlang]
y = 1;
\end{lstlisting} &
\begin{lstlisting}[xleftmargin=3pt,style=customlang]
a1 = x; //1
isync;
b1 = y; //0
\end{lstlisting} &
\begin{lstlisting}[xleftmargin=3pt,style=customlang]
b2 = y; //1
isync;
a2 = x; //0
\end{lstlisting}
    \end{tabular}
    \caption{IRIW: forbidden by multi-copy-atomic models, permitted by non-multi-copy-atomic models}
    \label{iriw}
\end{figure}

In the program of Fig.\ \ref{iriw}, the isync (instruction synchronization) ensures that the reading threads satisfy the reads in program order. The reading threads have no choice but to satisfy their reads from the propagation unit. The execution of the first reading thread implies that the write to $x$ was flushed before the write to $y$; the execution of the second reading thread implies the opposite. This contradictory behavior cannot be observed on multi-copy-atomic models. 

However, these independent writes, if placed in a non-trivial propagation unit of a non-multi-copy-atomic model, can be propagated in different orders to the reading threads, and permit the observation illustrated in Fig.\ \ref{iriw}.

\subsubsection{RMO}
The Relaxed Memory Order (RMO) model of SPARC(v9) enforces processor self-consistency, dependencies, and has a set of fences, as specified in \cite[Appendix D]{SPARCv9}. Interestingly, it does \emph{not} enforce SC-per-location, making it the only model to allow the behavior illustrated in Fig.\ \ref{loadloadhazard}: here, in the absence of explicit synchronization, the program-order-later load into $b$ is allowed to overtake the load into $a$. To quote the documentation itself \cite[\S 8.4.4.1]{SPARCv9}.
\begin{quote}
Relaxed Memory Order places no ordering constraints on memory references beyond those required for processor self-consistency. When ordering is required, it must be provided explicitly in the programs using MEMBAR instructions.
\end{quote}

\begin{figure}[h]
    \setlength{\tabcolsep}{8pt}
    \centering 
    
    \begin{tabular}{c||c}
\begin{lstlisting}[xleftmargin=3pt,style=customlang]
a = x; //1
b = x; //0
\end{lstlisting} & 
\begin{lstlisting}[xleftmargin=3pt,style=customlang]
x = 1;
\end{lstlisting} 
    \end{tabular}
    \caption{Loads from the same location racing: permitted only by RMO, this particular violation of SC-per-location is forbidden by all other models we consider.}
    \label{loadloadhazard}
\end{figure}

\subsubsection{ARMv8}
This model is excellently documented in \cite{armedcats,ARMv8}. It enforces SC-per-location, forbidding the behavior of Fig.\ \ref{loadloadhazard}. Overtaking in the buffer is constrained by SC-per-location and \emph{preserved program order}. We also recall that ARMv8 has introduced load-acquire and store-release instructions in its instruction set; see \S Fences of Sec.\ \ref{buffer}.

In contemporary descriptions of memory models, dependencies are captured by the notion of {preserved program order}. In descriptions of multi-copy-atomic models, it makes no technical difference when preserved program order also subsumes fences, but if there is no multi-copy-atomicity, fences have additional synchronization duties in the propagation unit, and must be distinguished.

\subsubsection{PSO}
The Partial Store Order (PSO) model of SPARC(v9) strengthens RMO by forbidding reads from being overtaken. Our abstract executing machine implements PSO by inserting a (load-load and load-store) memory barrier after every read.

\subsubsection{TSO}
The Total Store Order (TSO) model of SPARC(v9) and x86 strengthens PSO by further forbidding writes from racing. Our abstract executing machine implements TSO by inserting a (load-load and load-store) memory barrier after every read, as well as a (store-store) memory barrier after every store.\footnote{The official documentation places the barrier after, but it can also be placed before to have the effect of allowing only stores to be buffered in a disciplined queue.} We observe that by construction, loads in TSO take acquire semantics, and stores take release semantics. Indeed, TSO is formally shown stronger than SRA \cite{LahavGV16}.

\subsection{Non-Multi-Copy-Atomic Models}
We now consider models where weak behavior can also be attributed to delays in the message propagation unit. In all models except WRA and PRAM/FIFO Consistency, the partial order of propagation enforces per-location coherence, i.e.\ the set of writes to the same location is totally ordered (see Fig.\ \ref{per-loc-co}).

\begin{figure}[h]
    \setlength{\tabcolsep}{8pt}
    \centering 
    
    \begin{tabular}{c||c||c||c}
\begin{lstlisting}[xleftmargin=3pt,style=customlang]
x = 1; 
\end{lstlisting} & 
\begin{lstlisting}[xleftmargin=3pt,style=customlang]
x = 2; 
\end{lstlisting} & 
\begin{lstlisting}[xleftmargin=3pt,style=customlang]
a1 = x; //1
isync;
a2 = x; //2
\end{lstlisting} &
\begin{lstlisting}[xleftmargin=3pt,style=customlang]
a1 = x; //2
isync;
a2 = x; //1
\end{lstlisting} 
    \end{tabular}
    \caption{Models that enforce per-location-coherence forbid this outcome, as the writes to $x$ must be totally ordered, and earlier writes must be rendered redundant upon reading later ones. Both WRA and FIFO, however, allow this outcome.}
    \label{per-loc-co}
\end{figure}

\subsubsection{POWER}
We refer to the model of \cite{AlglaveMT14}, which was validated by extensive testing on hardware. The overtaking in the transaction buffer is constrained by preserved program order (corresponding to dependencies) and fences. Fences have further synchronization duties: they are responsible for creating causal dependencies to refine the propagation (partial) order \cite[Fig.\ 18]{AlglaveMT14}.

We illustrate the cumulativity of fences through two examples. The cumulativity of the lightweight fence is shown in Fig.\ \ref{mp}. It ensures that the write to $y$ is ordered after that to $x$. Thus, upon reading the write to $y$, the initial write to $x$ is rendered redundant, forbidding the illustrated outcome.
\begin{figure}[h]
    \setlength{\tabcolsep}{8pt}
    \centering 
    
    \begin{tabular}{c||c}
\begin{lstlisting}[xleftmargin=3pt,style=customlang]
x = 1; 
lwsync;
y = 1; 
\end{lstlisting} & 
\begin{lstlisting}[xleftmargin=3pt,style=customlang]
b = y; //1
isync;
a = x; //0
\end{lstlisting} 
    \end{tabular}
    \caption{Message passing. The annotated outcome is forbidden in POWER as well as in WRA, as the assumption of per-location coherence is not used.}
    \label{mp}
\end{figure}

The full fence (sync) provides even stronger synchronization. To illustrate it, we revisit the IRIW example and make a slight modification (see Fig.\ \ref{iriwsync}): the isync instructions in the readers are replaced by sync fences, which are flushed to the propagation unit. These fences must be totally ordered in the buffer, in the order in which they are flushed. Without loss of generality, we assume that the first reader flushes its fence first. This full fence is ordered after the write $x = 1$. When the second reader flushes its fence, it is ordered after the first fence, and by transitivity, also after the write $x = 1$. Recall that fences are implemented as release-acquire RMWs to an otherwise unused location: thus, this flush renders the initial write to $x$ as redundant to the second reader. The behavior illustrated in Fig.\ \ref{iriwsync} is therefore forbidden.

\begin{figure}[h]
    \setlength{\tabcolsep}{8pt}
    \centering 
    
    \begin{tabular}{c||c||c||c}
\begin{lstlisting}[xleftmargin=3pt,style=customlang]
x = 1;
\end{lstlisting} & 
\begin{lstlisting}[xleftmargin=3pt,style=customlang]
y = 1;
\end{lstlisting} &
\begin{lstlisting}[xleftmargin=3pt,style=customlang]
a1 = x; //1
sync;
b1 = y; //0
\end{lstlisting} &
\begin{lstlisting}[xleftmargin=3pt,style=customlang]
b2 = y; //1
sync;
a2 = x; //0
\end{lstlisting}
    \end{tabular}
    \caption{IRIW+sync: forbidden by POWER as well as by WRA}
    \label{iriwsync}
\end{figure}

\subsubsection{SRA}
SRA stands for Strong Release-Acquire. It was identified in \cite[Section 4.7]{AlglaveMT14} as a means to make the propagation under RA partially ordered, and formally developed in \cite{LahavGV16}. As \cite{LahavGV16} proves, the standard compilation scheme of \cite{cpppower1,cpppower2} (place lwsync before every write, place a fake control dependency and isync after every read) to POWER results in SRA. This is also how we compile SRA programs to our executing machine. 

Under SRA, reads thus have a \emph{clogging} effect on the buffer, and writes by a process may not race. Such disciplined store buffering is subsumed by the propagation unit. In our framework, we can thus assume that the weak behavior of SRA is entirely attributed to the propagation unit, and transaction buffers are always empty, i.e.\ the PC and ST coincide, and transactions are instantly flushed or satisfied.

To summarize: transactions are satisfied from, or flushed to the propagation unit in program order, the model is causally consistent (because writes accumulate the observations made by the author), and incoming writes are always placed as maximal elements of the total coherence order of the location.

\subsubsection{PSI}
Parallel Snapshot Isolation (PSI) is a consistency model used in databases and distributed systems that offers scalability and availability in large-scale geo-replicated systems \cite{psi}. Following \cite[Section 3.1]{LahavXRA}, we consider the restriction of PSI to single-instruction transactions. If all writes are replaced by RMWs, SRA precisely captures PSI.

\subsubsection{WRA}
WRA, as formulated in \cite{LahavXRA}, stands for Weak Release Acquire, and is a causally consistent model. WRA (without RMWs) is equivalent to a basic causal consistency model called CC in \cite{BouajjaniEGH17}, when CC is applied to the standard sequential specification of a key- value store supporting read and write operations.

In order to run WRA, our executing machine must not insist on a total per-location coherence order in its propagation unit: it must simply require that writes by the same process to the same location be totally ordered. The compilation scheme from source code is the same as that for SRA: lwsync before writes, and control with isync after reads.

RMWs only require weak atomicity: no two RMWs may read from the same write. Nevertheless, RMWs can still be used to implement a full fence \cite[Example 3.9]{LahavXRA}. Notice that we used this observation in the iriw+sync litmus test of Fig.\ \ref{iriwsync}.

Consequently, WRA can be summarized similarly to SRA (differences \emph{emphasized}): transactions are satisfied from, or flushed to the propagation unit in program order, the model is causally consistent (because writes accumulate the observations made by the author), and incoming writes are always placed as maximal elements of the total \emph{po-loc} order of the writes \emph{made to the location by the writing process}.

Interestingly, WRA allows a particular form of store-forwarding, illustrated in Fig.\ \ref{sf}. This is because there is no mechanism in the propagation unit to order the writes by different processes, and reading one does not render the other redundant.
\begin{figure}[h]
    \setlength{\tabcolsep}{8pt}
    \centering 
    
    \begin{tabular}{c||c}
\begin{lstlisting}[xleftmargin=3pt,style=customlang]
x = 1; 
a = x; //2
b = x; //1
\end{lstlisting} & 
\begin{lstlisting}[xleftmargin=3pt,style=customlang]
x = 2;
\end{lstlisting} 
    \end{tabular}
    \caption{Store forwarding. Synchronization instructions inserted at compile time are left implicit. This outcome is allowed by WRA.}
    \label{sf}
\end{figure}

\subsubsection{FIFO Consistency}
Finally, for completeness\footnote{(and to atone for the misrepresentation in \cite{originalCAV})}, we discuss FIFO Consistency \cite[Section 3]{fifosourceOG} (see also \cite[Section 3]{fifosourcesec}). We do not treat RMWs for lack of coherence in the model (this lack of coherence is by design, since the model is most often intended for distributed systems). Our framework needs adaptations to accommodate FIFO, but the key idea of a partially ordered propagation unit remains valid.

The FIFO protocol is as follows: each process maintains a local copy of the shared memory. A read fetches the value held in the local copy. A write overwrites a local copy, and broadcasts a message. The local copy may be overwritten by a broadcast from another process at any time. It is guaranteed that messages from any process will overwrite any local buffer in program order.

This means that FIFO can ``pass'' messages (the outcome of Fig.\ \ref{mp} is forbidden under FIFO when the synchronizing instructions are edited appropriately), but is unable to ``relay'' messages: see Fig.\ \ref{mrelay}. This is because under FIFO Consistency, the writes of the first and second processes are propagated \emph{independently} to the third process.

\begin{figure}[h]
    \setlength{\tabcolsep}{8pt}
    \centering 
    
    \begin{tabular}{c||c||c}
\begin{lstlisting}[xleftmargin=3pt,style=customlang]
x = 1; 
\end{lstlisting} & 
\begin{lstlisting}[xleftmargin=3pt,style=customlang]
a = x; //1
y = 1; 
\end{lstlisting} &
\begin{lstlisting}[xleftmargin=3pt,style=customlang]
b = y; //1
c = x; //0
\end{lstlisting}
    \end{tabular}
    \caption{Message relay. The annotated outcome is forbidden if the program is compiled as WRA, but allowed if the program is compiled as FIFO.}
    \label{mrelay}
\end{figure}

It turns out that FIFO is incomparable to the models we have seen: the protocol is incompatible with store buffering, as the litmus test in Fig.\ \ref{noSB} shows. The annotated outcome is observed under TSO if the second process reads $x$ to $a_1$ and $y$ to $a_2$ right between the flushing of $x = 1$ and of $y = 1$, reads from $y$ to $a_3$ after the flushing of $y = 2$, and forwards $x = 3$ from the buffer to satisfy the read of $x$ to $a_4$.

If the program were running under the FIFO protocol, then upon the execution of the first three instructions of the second process, we know that the write $x = 1$ had been conveyed to the local copy, it was subsequently overwritten by $x = 3$, but the write $y = 1$ is yet to be conveyed because the local copy still contains the initial value. The next read implies, by the FIFO property that both $y=2$ and $x=2$ have been conveyed to the local copy, the latter \emph{overwriting} $x = 3$. Hence, it is subsequently impossible to read $x = 3$ into $a_4$.

\begin{figure}[h]
    \setlength{\tabcolsep}{8pt}
    \centering 
    
    \begin{tabular}{c||c}
\begin{lstlisting}[xleftmargin=3pt,style=customlang]
x = 1; 
y = 1;
x = 2;
y = 2;
\end{lstlisting} & 
\begin{lstlisting}[xleftmargin=3pt,style=customlang]
a1 = x; //1
x = 3;
a2 = y; //0
a3 = y; //2
a4 = x; //3
\end{lstlisting} 
    \end{tabular}
    \caption{Store-buffering behavior allowed by TSO but forbidden by FIFO (appropriate barriers implicit)}
    \label{noSB}
\end{figure}

We now express the FIFO protocol with our framework. We keep the compilation of FIFO programs to our executing machine abstract, as it requires a non-standard instruction set for the required synchronization. Instead, we simply declare the policy of the transaction buffers, and structure and maintenance of the partially ordered message propagation unit. 

The transaction buffers are always assumed to be empty, i.e. \emph{all} instructions have a clogging effect. The program counter and speculation tracker always coincide, reads are satisfied immediately, writes are flushed immediately.

The key design choice of FIFO is that there is no global synchronization. However, \emph{all} writes by the same process are totally ordered, and the total order is the same as the program order. Thus, the partial order of the propagation unit consists of one chain per process. However, elements from distinct chains are always incomparable. 

To capture FIFO, we require that for each process $p$ and location $x$, \emph{exactly} one message to $x$ be both seen by and not redundant to $p$. Thus, if a message is marked as seen to $p$ during an update step or a write, then other non-redundant messages that were earlier marked seen become redundant. The ensuing garbage collection then ensures that po-earlier writes to that variable also become redundant.

\begin{description}
\item [FIFO Update] Choose a process $p$, location $x$, and message $v_0$ to $x$ that is not redundant to $p$. Mark $v_0$ as seen by $p$. Other non-redundant (to $p$) messages to $x$ that are seen by $p$ get marked redundant, and garbage collection ensues. Mark the following set of messages as redundant to $p$: \\
$\{u: \exists v. ~ (u \text{ and } v \text{ write to the same variable}) \land (u < v \le v_0) \}$. \\
Then, mark messages $v$ such that $v \le v_0$ and $v$ is not redundant to $p$ as seen by $p$. If $v$ is a message to $y$, other non-redundant (to $p$) messages to $y$ that are seen by $p$ get marked redundant, and garbage collection ensues.
\item [FIFO Read] Return the value held in the unique non-redundant message to $x$ seen by $p$.
\end{description}

%% file: inherit.tex

\section{Related Work}
\label{related}

\subsection{Fairness}
Only recently has fairness for weak memory
started receiving increasing attention.
The work closest to ours is by \cite{DBLP:conf/esop/AbdullaAAGK22},
who formulate a probabilistic extension for the
Total Store Order (TSO) memory model
and show decidability results for associated verification problems.
Our treatment of fairness is richer, as we relate same probabilistic fairness with two alternate logical fairness definitions.
Similar proof techniques notwithstanding, our verification results are also more general, thanks to the development of a uniform framework that applies to a landscape of models.
%
%
The authors of \cite{strongcohdef} develop a novel formulation of fairness 
as a declarative property of event structures. This notion informally translates to ``Each message is eventually propagated.''
We forego axiomatic elegance to motivate and develop stronger practical notions of fairness in our quest to verify liveness.

\subsection{Framework}
As acknowledged in the Introduction, for formulating our framework, we draw inspiration from the operational model in \cite[Section 7]{AlglaveMT14}, which is itself adapted from the authors' line of previous work, most notably \cite{genericmodel}. We formulate our models in the documentation-style of \cite{armedcats,ARMv8}. However, we distinguish ourselves by choosing to enforce a minimal set of axioms by default, and putting simple, garbage-collection-friendly data structures like queues and a partial order at the forefront. This formulation easily keeps track of only the information that is relevant to the future, and lends itself to memory fairness. To the best of our knowledge, our work presents a unique combination of modeling, fairness notions, and verification of $\omega$-regular linear temporal properties.

On the modeling front, 
the ability to specify memory model semantics
as first-order constraints over the program-order, 
reads-from relation, and per-location coherence order
have led to elegant declarative frameworks based on event structures
\cite{Alglave2012AFH,AlglaveMT14,Chakraborty2019GroundingTR,Jeffrey2016OnTA}.
There are also approaches that, instead of natively characterizing semantics,
prescribe constraints on their ISA-level behaviors
in terms of program transformations \cite{Lahav2016ExplainingRM}.
On the operational front, there have been works that 
model individual memory models \cite{NienhuisMS16,SSONM2010} 
and clusters of similar models \cite{KangHLVD17,LahavGV16},
however we are not aware of any operational modeling framework that
encompasses as wide a range of models as we do.
The operationalization in \cite{Boudol2009RelaxedMM} uses
write buffers which resemble our channels, however, their operationalization
too focuses on a specific semantics.

\section{Future Work and Perspective}
\label{conclusion}
\subsection{Future Work}
There are multiple interesting directions for future work.
It is interesting to mix transition fairness with probabilistic fairness, i.e., use the former to solve scheduler non-determinism and the latter to resolve memory non-determinism, leading to (infinite-state) Markov Decision Process model.
Along these lines, we can also consider synthesis problems based on $2\frac12$-games.
To solve such game problems, we could extend the framework of Decisive Markov chains that have been developed for probabilistic and game theoretic problems over infinite-state systems \cite{AbdullaHM07}
A natural next step is developing efficient algorithms for proving liveness properties for programs running on weak memory models.
In particular, since \cite{originalCAV} reduce the verification of termination and repeated control state reachability to simple reachability, there is high hope one can develop CEGAR frameworks relying both on over-approximations, such as predicate abstraction, and under-approximations such as bounded context-switching \cite{DBLP:conf/tacas/QadeerR05} and stateless model checking \cite{AbdullaAJS17,Kokologiannakis18}.

\subsection{Perspective}
Leveraging techniques developed over the years by the program verification community, and using them to solve research problems in programming languages, architectures, databases, etc., has substantial potential added value. 
Although it requires a deep understanding of program behaviors running on such platforms, we believe it is about finding the right concepts, combining them correctly, and then applying the existing rich set of program verification techniques, albeit in a non-trivial manner.
The current paper is a case in point.
Here, we have used a combination of techniques developed for reactive systems \cite{DBLP:journals/fmsd/KestenPRS06}, methods for the analysis of infinite-state systems \cite{AbdullaHM07}, and semantical models developed for weak memory models \cite{DBLP:conf/popl/AtigBBM10,KangHLVD17,LahavXRA,LahavGV16} to obtain, for the first time, a framework for the systematic analysis of liveness properties under weak memory models.